# Atomic Biology, Electrostatics, and Ionic Channels

**RS Eisenberg**

Department of Molecular Biophysics and Physiology

Rush Medical College

1750 West Harrison Street

Chicago IL 60612

USA

# Table of Contents

## MOLECULAR BIOLOGY

The identification and manipulation of the molecules of life is one of the triumphs of man, putting biological systems under his control, embedding them into his technology, allowing the control of many of life's functions. In the next few years some of the dreams of centuries will become realities, as we learn to control disease and our biological life cycle in ways that only poets and novelists have imagined, by controlling proteins.

The triumph of this molecular biology is based on the identification and manipulation of proteins, and of their blueprints, DNA and some kinds of RNA. The technology of molecular biology manufactures these blueprints and thereby allows manipulation and bulk production of the encoded protein. This impressive technology is possible because so many biological functions can be controlled, or influenced, simply by the presence or absence of specific proteins in sufficient amounts. Evolution has used this simple method of control probably because it is easier to make adaptations this way, depending on the mutation and selection of single genes, rather than in other ways that depend on changes in many genes, and therefore occur with much lower probability, if the mutation of each of the genes is independent, as traditionally assumed.

If evolution has used this simple mechanism to control many systems, then we can use it too; we can take advantage of this simplicity and govern those systems ourselves, even if we do not understand them. A car can easily be controlled by a driver who knows little of how it works. Similarly, a metabolic pathway governed by a single enzyme can be controlled technologically long before the pathway is known or understood. Molecular biology, as presently practiced, allows the control of biological functions that depend on the concentration of one protein. It will allow less control in systems that depend on the presence and properties of many proteins, although let us hope these are few and far between (for the sake of our own health). The quip is that it will be possible to grow a nervous system long before we understand it![1]

---

[1] because the growth is determined (in all likelihood) by hundreds or a few thousands of genes whereas the function of the mammalian nervous system is determined (potentially) by its some $10^{13}$ cells and their multitude of connections. Many of those cells and connections have to be investigated individually to determine how the nervous system functions. The problem is simpler in the much smaller nervous systems of invertebrates, but it is

Molecular biology is also a useful tool for analyzing the function of a protein. Molecular genetics allows the rational manipulation of a protein's primary (and, perhaps someday, secondary and tertiary) structure, an experimental technique of the greatest importance. But the simple replacement of one amino acid by another (e.g., in site-directed mutagenesis and its congener techniques) cannot tell how the protein works, how that amino acid influences function any more than the simple replacement of the parts of a machine, an automobile for example, will tell us very much about how it works. Replacement of amino acids may show that one controls a function of an enzyme, just as replacement of the spark plugs will show that they (along with many other parts of the engine) control ignition. But the identity of the other parts will remain unknown, and so will their interactions with the spark plug. And how that amino acid (or spark plug) works will often remain a mystery, as well.

Changing one amino acid cannot in general have a specific effect on the physical properties of a protein. There are too many amino acids in a protein; any one amino acid is expected to have subtle, diffuse and unimportant overall effects on physical properties. A particular amino acid residue may, nonetheless, be able to control function of an entire protein for essentially evolutionary, not physical reasons, even if most individual amino acids in the protein have a small effect on the properties or function of the whole protein. After all, the fundamental mechanism of evolutionary change is the substitution of one amino acid by another. Proteins evolve by randomly changing amino acids until one produces a function with selective advantage. Changes in each amino acid are (more or less) independent random events. Thus, the probability of multiple useful changes is very very much less than the probability of a single useful change. For this reason, many functionally significant properties of proteins are likely to be controlled by a particular amino acid, the amino acid found by evolution to produce the function it needed!

But saying that an amino acid controls a function says nothing about how it controls that function. A light switch controls the light; and knowing where the switch is and how to use it is of immense practical importance. But locating the light switch tells nothing about how the light

---

not clear whether the distinctive behaviors of mammals—that we (anthropocentrically) call "intelligent life"— occur in invertebrates.

works, about how the switch works, or about how they are coupled together. The light might be incandescent, fluorescent, or whatever; the switch might be mechanical, electronic, or something else; the coupling might be electrical, electromagnetic (e.g., an infrared remote control), or even mechanical. And the switch might be inches, feet, or yards (indeed, miles, in special cases) away from the light.

Controlling protein function to improve our technology is one thing, of great practical importance; but understanding the function, to extend our knowledge, is another. To understand we need a specific physical model with atomic resolution, I believe. The techniques of molecular biology have too coarse resolution to analyze or describe the physical mechanisms underlying protein function because the role of atoms is an important part of the physical basis of protein function.

### ATOMIC BIOLOGY

I believe an atomic biology is needed to supplement present day molecular biology, if we are to design and understand proteins, as well as define, make, and use them.

The challenge of understanding proteins in general—the grand challenge of atomic biology—is a formidable one, one I fear that is beyond us for now, certainly beyond me. Most significant proteins are enzymes that catalyze chemical reactions. Such reactions occur in solutions, in condensed phases with little space between atoms; they almost always involve changes in covalent bonds. The underlying physics of enzymes is the physics of chemical reactions in solutions, of quantum chemistry in condensed phases. Here experimental and theoretical problems are too formidable to solve, even for unstructured chemical reactions in homogeneous bulk solutions. The chemical reactions of enzymes would be just as difficult to understand, even if the reactions were spatially homogeneous like chemical reactions in bulk solutions. But most enzymes have a specific three-dimensional structure which is anything but homogeneous. Indeed, enzymes direct structured chemical reactions, achieving many of their remarkable properties this way. It is unlikely that the structured chemical reactions typical of enzymes can be understood until chemists better understand chemical reactions in the condensed phase in general; until experimental techniques allow the visualization of the directed structural changes of enzymes and their substrates; until experimental techniques resolve the time course of those structures and their transformations. This science is certainly beyond our grasp, and will even stay out of our reach for some time to come, I fear.

The promise of atomic biology then seems a distant trumpet compared to the clarion call of molecular biology. Why even try to answer such a feeble summons? The reason is that a trumpet's sound becomes loud when focused on an appropriate object, even loud enough to bring walls tumbling down, so they say. The promise of atomic resolution may be reached if we focus on a class of proteins simpler than enzymes, on ionic channels (Hille, 1992).

Ionic channels are proteins that catalyze (Eisenberg, 1990) the movement of solutes from one place to another, from outside a cell to inside a cell, or outside an organelle to inside an organelle. Channels do not catalyze changes in covalent bonds and so understanding their function does not require understanding the (quantum) theory of chemical reactions in solutions.[2] Yet the biological importance of these channel proteins is hard to overstate. They are nearly as important as enzymes. They sit in membranes in series with the metabolic machinery of cells. Just as security in an airport is best provided by controlling access, so is control of cellular functions easily provided by specialized gates, selectively filtering entering ions and metabolites, thereby restricting access to the interior of cells and organelles. Channels control access to the cell interior the way doors control access to buildings. Channels provide control by regulating the diffusion and drift—the electrodiffusion—of ions across membranes. In this way, they control a substantial fraction of the functions of life. Channels control the electrical function of cells, for example, signaling in the nervous system. They control muscle function by coordinating contraction, for example, of the ventricle of the heart, thereby allowing it to pump blood[3]; they are involved in the transport and secretion of most substances in cells and organs, including uncharged molecules. Channels are ideally placed to control the life of the cell, and the large number of drugs that act on channels allow physicians and scientists useful control of cells of real importance to our daily lives.

---

[2]Quantum effects are likely to be important in understanding (1) interactions of ions with aromatic residues in the channel wall (Dougherty and Kumpf, 1993) and (2) the interactions of ions with nonpolar, but polarizable hydrogens in the channel wall (D.P. Chen, C. Lee, and H. Chen, personal communication).

[3]without co-ordination, the ventricle contracts pointlessly in impotent fibrillation and is unable to move blood anywhere useful.

Ionic channels are important, yet they have a simple structure that controls function in a simple way. Channels have an aqueous pore down their middle, a pore we might call a channel—a channel within a channel— if that were not so confusing.[4] In this pore ions move following laws of electrodiffusion much as they do in solution. In classical work on channels and membranes (before 1981, say), currents through individual molecules were not resolved. The currents recorded from macroscopic amounts of membrane were usually interpreted as flows through a fixed area of homogeneous membrane. We now know that the area through which macroscopic currents flow is anything but fixed. The number of open channels varies with a wide range of conditions and thus so does the area for current flow. Channels open and close in response to stimuli, and when they are open, molecules move by electrodiffusion.

The classical interpretation of macroscopic currents has been discarded in the face of these new highly resolved measurements. Many of the phenomena attributed to "the membrane" in classical theories arise from changes in the number of open channels, i.e., in changes in the area available for current flow. Such gating phenomena are not the subject of this paper (except peripherally); rather we concentrate on the properties of the channel protein, and of the channel's pore, once it is open. Classical theories are still relevant, however, even if classical descriptions of data are not. The ***classical theories now describe the flow of ions through one open channel***, not through a membrane. Flowing through a channel, each ion follows a stochastic trajectory starting in the surrounding baths. Collectively, the drift component of the random trajectories of individual ions produce a diffusive flux driven by the (bath) concentrations of permeating ions as described (in the mean) by classical theories.

---

[4]'Channel' in dictionaries usually means "a passage through which something can be directed ... a tubular passage ... a conduit" (Soukhanov, 1992). Biologists, however, sensibly identify a channel as ***the protein***, not the pore, perhaps because biologists are more interested in the protein than its contents, just as a steelworker is more interested in the metal that makes his 'channel' (a rolled metal bar) than the air it encloses.

The opening and closing of channels—called 'gating' in the literature of channology—is not a simple process. Gating determines much of the biological function of channels and is thought to be a conformation change, a substantial change in the position of many of the atoms of the channel protein. If such, it is not immediately amenable to the high resolution analysis of atomic biology. Conformation changes are usually observed only indirectly if at all. When direct structural methods do resolve movements in proteins, those movements are often of (artificially or naturally) 'tagged' parts of the protein that may not participate in biological function in a central way and so may not follow a simple, reasonably robust set of phenomenological laws.[5] Thus, it is not surprising that the physics of (functionally relevant) conformation changes is not known; indeed, we do not know if there are any general principles at all that govern them. Conformation changes are described and named in molecular biology, but they are not often measured directly or analyzed. The physical forces and atomic interactions that produce these changes in protein shape are hardly ever known. Conformation changes of channels cannot yet be analyzed with a physical theory at atomic resolution.

For this reason, if we insist on the agenda of atomic biology—the agenda of physical understanding of biological function at atomic resolution—we restrict analysis of channels to their pores, to the already open channel, at least to begin with.[6] We lose much interesting biology, including the analysis of traditional mechanisms of gating. In return we gain simplicity, hoping that permeation through the open channel is a humble enough process occurring in a well enough defined structure that we can reduce it to a physical problem with atomic resolution.

Perhaps the open channel will show what atomic biology can do.

### ELECTRODIFFUSION IN THE OPEN CHANNEL

Movement of ions through channels is driven by diffusion in a concentration gradient and drift in an electric field. Both diffusion and drift are significant in every channel, in nearly every situation; neither can be approximated away, because both are important in determining biological function. The driving force for both will appear in any consistent[7] model of a channel relevant to biological function or experimental data.

---

[5]See the discussion of **Simplicity, Evolution and Natural Function** later in this paper

[6]However, permanently open channels of one conformation can gate (at least in principle) by changing the shape of their electric field (see **Gating by Field Switching** in this paper, and Chen and Eisenberg, 1993a,b).

***Drift in the Electric Field***. The electric field is the driving force for this motion and the electric field is determined by the types and distribution of charges, all important, all of which will appear in an atomic theory.

(1) ***Charge in the channel's pore.*** The contents of the channel are one kind of charge, namely, the ions in the channel's pore. These have much the same concentration as the permanent protein charge on the wall of the channel (because they more or less neutralize the protein charge). They serve to shield or screen the permanent charge of the protein and so cannot be neglected. Indeed, the contents of the channel will change if the flux through the channel is changed by experimental manipulations of (trans-membrane) potential or concentrations (in the bath). The charged contents of the channel have often been neglected in treatments of ion movement through channels, although they have not been neglected in the analogous problem of hole and electron conduction in semiconductors (Jerome, 1994; Shur, 1990). There, the contents of the semiconductor 'channel' can change the electric field (in, for example, a thyristor, Sze, 1981; Taylor, 1987) and make the semiconductor device turn ON or OFF.

Almost any molecule in a channel's pore has significant charge. Water is a highly polar molecule: each of its atoms has significant charge. Even nonelectrolytes soluble in water have a significant distribution of electric charge. That is why they dissolve in a polar solvent! In particular, biological nonelectrolytes like glucose have nonuniform distributions of electric charge in their chemical bonds. Most of those bonds are quite polar, and partial electric charge on most of the atoms of glucose is substantial, of the same order of magnitude as the charge on an ion. The partial charges on glucose sum to neutrality, to be sure—so glucose is a nonelectrolyte—but they are separated by angstrom distances, roughly the same as the distance between an ion and the channel wall. The partial charges within nonelectrolytes cannot be neglected. Even a nonelectrolyte like glucose is a collection of charge. And it is better to describe that charge directly as a charge density than to use a few terms (typically, monopole and dipole, sometimes quadrupole) in a multipole expansion, which, all too often, takes hundreds of terms to converge.

---

[7]precisely defined later in the text, on page 29.

> Close to a (bounded uniform) distribution of charge, where the electric field is strong and the charge interacts most significantly with ions, proteins, etc., the multipole expansion (Panofsky & Phillips, 1962, p. 82-87; Hobson, 1965) has an $n^{th}$ term that behaves (in magnitude) like $\left(R/r\right)^n \left(1/n\right)$ where $r$ is the distance from the center of charge to the observation point and $R$ is the distance to the charge itself $\left(R/r\right) \leq 1$. Thus, the rate of convergence is proportional to $R/r$. This is indeed weak convergence, particularly just where the field is largest and most important, near the charge distribution, when $r \cong R$. Close to the charge distribution, hundreds of terms are needed to reach even 10% accuracy.

(2) ***Protein's permanent charge.*** The channel wall, the surface of the channel protein, is also a collection of charges. The channel wall is a distribution of permanent protein charge determined by the structure of the protein, a charge that does not change unless the structure of the protein changes.

***Charge resides in nearly all the bonds of the protein, not just in those with formal charges***. Most of the bonds of a protein are polar and the partial electric charge on most of the atoms of a protein is substantial, of the same order of magnitude as the charge on an ion or an atom of the protein with a full formal charge (Schulz and Shirmer, 1979; Gilson and Honig, 1988; Creighton, 1983). The partial charges are separated by angstrom distances, the same distance between an ion and the channel wall, and so the partial charges are of great significance and cannot be neglected. A protein is a collection of charges.

Because a protein is a collection of permanent charges, its interaction with substrates is quite different from that described by usual enzyme kinetics (e.g., Michaelis-Menten kinetics, e.g., Fersht, 1985; Walsh, 1979). Indeed, the interactions are rather different from those implied when a chemical reaction, or ion permeation, is described as the movement of a particle in a "potential of mean force".

If the surface of a protein is described as a potential of mean force (one of the mean-force potentials described by Hill, 1956) that potential of mean force is usually taken as independent of the concentration of particles crossing it. Indeed, the potential structure (i.e., electric field) is usually treated as a constant, independent of concentration of solutes in the bath, independent of trans-membrane potential (if the protein is a channel), or electrochemical potential (if the protein is an enzyme), independent of the permeating particle. However, if the protein is a surface of permanent charge, the potential of mean force cannot be a constant independent of these variables because anything that changes the distribution of charge in the system will change the shape of the potential of mean force even if the conformation of the protein remains strictly fixed, as we will discuss in some detail later in this paper.

Under one set of experimental conditions, the surface of a protein can be represented as a distribution of charge, or the ***corresponding*** distribution of potential; the two representations are strictly equivalent, for that set of conditions, ***but no other***. If concentrations are changed, or if trans-membrane potential is changed, the distribution of net charge is changed—even if the conformation of the protein, and thus the distribution of permanent charge, remain strictly fixed—and the potential of mean force cannot remain the same, unless charge is supplied or withdrawn from the protein. That cannot happen (since the protein is an isolated object, not connected to a battery in these systems). The permanent charge stays the same as experimental conditions change. But the net charge does change and so does the potential. Thus, on the most fundamental grounds, the ***potential of mean force is a function of experimental conditions*** even if the structure of the protein does not change.

> The field equations have a unique solution if (for example) charge is specified on the boundary, that is to say if Neumann boundary conditions are imposed on the field in the first place. The solution to the Neumann problem then describes all the properties of the field—the distribution of potential and charge (etc.) everywhere, including at the boundary. (That is what 'unique solution' means!) The potentials (etc.) that are derived at the boundary from the unique solution to the Neumann problem are the ***only*** values compatible with the charges specified on the boundary in the first place. Anything that changes conditions, even if it leaves the boundary conditions (i.e., charges in the Neumann problem) untouched, will change the potential everywhere, including at the boundary. ***The relation between charge and potential on the boundary, and everywhere else, depends on the entire problem***.

> An analogous argument applies to any set of boundary conditions that imply a unique solution of the field equations, e.g., Dirichlet boundary conditions that specify the potential on the boundary or the rather complex boundary conditions necessary to describe a biological channel (Barcilon, 1992; Barcilon, Chen, and Eisenberg, 1992).

This dependence of the electric field and potential of mean force on experimental conditions has many implications for both channels and enzymes. Most models of both kinds of proteins seek to explain data taken under a range of experimental conditions, e.g., different concentrations of substrate or permeant ions, different potentials, and sometimes different ionic strengths. These models typically describe the protein as moving from state to state, following a simple rate law usually given by Eyring rate theory as a function of the potential of mean force on the surface of the protein (e.g., Hill, 1956; 1985; Hille, 1992). The rate constants in these rate laws are, of course, taken as independent of concentration of substrate—that is why they are called constants—and the dependence on potential is taken as exponential. But the potential of mean force on the surface of a protein or channel depends on the (average concentration of) substrate in the channel's pore or on the protein's surface. That substrate shields the permanent charge to a variable extent, depending in a nonlinear way on its concentration and the overall electric field. That shielding (or screening as it is sometimes called) changes the net charge (i.e., substrate charge plus protein charge) and so it changes the potential of mean force. Assuming that the potential of mean force is independent of substrate concentration is equivalent to assuming that the potential of mean force is independent of charge density; it cannot be true.[8]

Thus, traditional models of channels that fit properties measured under different conditions need to use different potentials of mean force under different conditions. Similarly, models of enzyme reactions need to use different potentials of mean force under different conditions (if they are to be consistent with the properties of the electric field) and rate 'constants' of such models are likely to vary with substrate concentration (and other experimental conditions), because the potential barriers that produce them vary. Of course, these different potentials of mean force under different conditions cannot just be assumed; they must be calculated from an electrostatic model of the protein, one that describes the protein's surface as a distribution of permanent charge, as (for example) does the PNP theory described later in this paper.

---

[8]although it may be a decent approximation in special cases. Of course, the validity of the approximation needs to be actually shown, not just assumed, in each special case.

The variation of the potential of mean force with experimental conditions has profound implications for traditional biochemical analysis of enzyme rates. In particular, the rate constants determined from a Michaelis-Menten analysis of enzyme kinetics will depend on substrate concentration, even if the conformation of the enzyme is independent of concentration. Perhaps some of the allosteric effects and complex kinetics found so often in traditional enzyme kinetics (Walsh, 1979) represent a failure in Michaelis-Menten theory, not a complex set of conformational changes of the enzyme.

(3) ***Distant charge and the trans-membrane potential***. Not only do channel proteins and permeating ions have electric charge, but so does the bathing solution they sit in, the bathing solution surrounding the cell membrane.

A trans-membrane potential is found across the outer membrane of nearly all cells, and across the membranes of many organelles, and that potential is used for biological function. For example, the cell membrane potential controls the opening and closing of many types of channels (so called voltage gated channels). The membrane potential is also one of the determinants of ionic motion in the channel's pore, once a channel is open. The electrical environment of channels is controlled by the entire cell: the lipid membrane in which channels sit supports an electrical potential produced by surface charge and by a variety of mechanisms in the cell and its membrane. In particular, the membrane potential is determined by the ionic conductance of the membranes (what we now recognize as the ensemble properties of single channels) and by the concentration gradients of ions.

Concentration gradients between cell inside and outside are maintained by specialized active systems that use energy ultimately derived from ATP. Some concentration gradients are maintained by the energy stored in other concentration gradients. For example, the substantial gradient of Ca$^{++}$ ion—some $10^{-3}$ M $(outside)\big/10^{-7}$ M $(inside)$ with typical membrane potential of -90 mV, inside negative—is maintained in part by a transporter protein (residing in the cell membrane) that exchanges Na$^+$ for Ca$^{++}$. The gradient for Ca$^{++}$ is also maintained by an active transport protein in the membrane—the Ca$^{++}$ pump—that hydrolyzes ATP and uses that energy to pump Ca$^{++}$ out of the cell, against its concentration and electrical gradient. The ultimate energy source for most concentration gradients (in most cells) is the ATP hydrolyzed by the Na/K ATPase as it pumps Na$^+$ and K$^+$ out of and into the cell.

The potential across a membrane is determined by processes far away, as is seen whenever a voltage-dependent channel is studied in its natural state, for example, when a Na$^+$ channel is studied in a nerve. There, the potential and the electric charge on a membrane ***millimeters*** away is the dominant determinant of the conformation of a protein! Because the membrane is such a good insulator—even without its myelin wrapper—the charge producing the membrane potential is spread over macroscopic distances. These distant charges certainly cannot be ignored: they help control the local electric field within the channel protein and are responsible for its natural function, the conduction of the nerve impulse. ***Consideration of nearby charges is not enough if a protein is sensitive to the electric field***.

(4) ***Interfacial charge***. The charge producing the membrane potential resides chiefly at the 'interfaces' of the membrane. Each of the several interfaces in the system has charge, both permanent (i.e., independent of the local electric field) and induced (i.e., dependent on the local electric field, absent when the field is zero). The induced charge in turn can depend linearly on the local electric field, and thus be traditionally described by a dielectric constant; or it can reside in the ionic atmosphere adjacent to the interface. Then it depends in a complex nonlinear way on electric field (and many other variables) as described approximately by Debye - Hückel theory in the equilibrium case or the PNP model in a continuum description of the nonequilibrium situation (Chen and Eisenberg, 1993a,b).

Consider the interface between the lipid and the bulk solution. Permanent charge is present there because the carbonyl bonds of the aqueous head groups have a substantial permanent polar charge and the nearby carboxylic acid moieties carry a permanent formal charge (at biological pH). Induced charge is present because of the ion atmosphere (the Debye-Hückel or Gouy-Chapman effect, depending on the choice of idealized geometry); because of the dielectric effects of the water and the lipid; and because of the trans-membrane potential. Charge near the lipid/solution interface can create a significant offset in the potential within the channel protein and the channel's pore (even though the interface can be distant from the channel) because the 'far-field' potential contributes an important term to the near field potential (Chen, Barcilon, and Eisenberg, 1992).

Modification of this far-field is likely to produce a dramatic change in channel properties simply by changing the electric field within the channel protein and its pore. Indeed, Chen and Eisenberg (1993a) have proposed this field coupling as the mechanism linking accessory proteins like receptors to ionic channels. They suggest that binding of an agonist to a receptor, or a subsequent conformation change of the receptor, or a chemical reaction involving a receptor (e.g., phosphorylation), is very likely to change the charge distribution of that receptor protein, thereby changing the electric field far from the channel itself. The far-field contributes a term to the potential within the channel even in the absence of specific chemical or allosteric interaction. The far-field would thus have a dramatic effect on gating, opening (or closing) a channel, and on permeation, modifying the rates of permeation. In this way *second messenger systems could be coupled to nearby channels by the electrostatic field*. Allosteric conformation changes of vague physical mechanism do not have to be invoked to explain the coupling of receptor to channel. The electric field can be the link, just as it is the link between the potential on the (distant) gate of a field-effect transistor and the potential in its conducting 'channel'.

(5) ***Ends of the Channel: Dehydration, Solvation, and Charge.*** Charge can also exist at the end of the channel, that is to say, at the interface between the pore and the bulk solution. Surface potentials might exist in that region that have direct effects on channel gating and permeation.

Large effects certainly occur at this interface between pore and solution because the solvation shell of an ion changes dramatically as it enters the channel. Such desolvation and resolvation effects may be dominant determinants of selectivity and flux coupling.

Ions in bulk solution are solvated by water. Ions in a channel's pore are solvated by the channel protein (mostly); they also interact with the two adjacent water molecules within the pore. As an ion enters a channel, it and the surrounding atoms undergo a large rearrangement, akin to a change in phase. The energies involved are huge (say 60 $kT$) compared to the energies important in permeation (say 6 $kT$). A few $kT$ has a large effect on permeation because rates of permeation tend to be exponential functions of energy barriers, with exponents given by the barrier height, in units of $kT$. Tiny mismatches between the energy of hydration (in the bulk material) and the energy of solvation (inside the channel) would have a huge effect on ion transport.

A phase boundary of this sort between channel and solution is likely to support complex phenomena, as it does in electrodes or gas/solid interfaces. Electrochemists describe the nonequilibrium properties of such interfaces by the Butler/Volmer equation (Bard and Faulkner, 1980); physical chemists describe the transport properties of the gas/solid interface by the closely related expressions of statistical rate theory (e.g., Ward, 1983; Ward and Elmoselhi, 1986). We should not be surprised if this interface in channels supports many complex phenomena and needs a description akin to other interfaces. The entry interface of channels may be responsible for the selectivity, and perhaps flux coupling (Chen and Eisenberg, 1993a, b) of channels and transporters.

The last interface we consider is between the pore and the channel protein where permanent charge and induced charge are present. The permanent charge has already been identified and analyzed as part of the contents of the channel, the permanent protein charge. Induced charge is found here and elsewhere in the channel.

(6) ***Induced charge*** is present at the interfaces, in the protein of the channel, the lipid of the membrane, and the water of the bulk solution. The local electric field displaces electrons in atoms, and other charge, from their 'rest' position (i.e., their position when the local electric field is zero). This induced charge is the dipolar polarization charge of dielectrics, described in traditional macroscopic theories by a polarization vector, an ideal dipole proportional to the local electric field (Panofsky and Phillips, 1962; Jackson, 1975; Stratton, 1941). Polarization charge in reality usually has complex time and field dependence (Fröhlich, 1958; Daniel, 1967; Böttcher, 1973; Hedvig, 1977; Pohl, 1978; Grant, Sheppard, and South, 1978; Macdonald, 1987; Eringen and Maugin, 1989; Mahan and Subbaswamy, 1990). It is always, however, induced and exists only when the electric field is non-zero. For some reason, the word 'polar' is also used to describe a chemical bond with a permanent charge distribution, independent of the electric field, making confusion easy: conventional usage might force one to say that "a polar bond contains no polarization charge".

Each type of charge must appear in a theory of channel permeation because each makes a significant contribution to the electric field that controls channel gating and open channel permeation. Otherwise, the theory will be inconsistent with the properties of matter or the electric field. Conventional macroscopic theory, in which electricity is a continuous fluid, allows easy mathematical description of all these types of charge and their interactions with each other and the electric field they produce. That is just the point of the PNP model (Chen and Eisenberg, 1993a,b), as we shall discuss later in this paper.

Electricity is not a fluid, however; it consists of discrete charges making stochastic movements and the discreteness of charge is clearly relevant in systems as small as ionic channels. The description of each type of charge, and their interactions with each other and the electric field they produce, is much harder in discrete consistent models. Indeed, it has not been done satisfactorily in bulk ionic solutions—say mixtures of $Na^+Cl^-$ and $K^+Cl^-$ and $H_2O$—let alone in protein solutions.

### *Diffusion through Channels*

Having described the electrical driving force, namely, the types of charge, we turn to the other driving force for flux through channels, concentration, or really activity (i.e., free energy per mole) for which the word 'concentration' is a widely used surrogate, an approximate stand-in.

A condensed phase like a protein in a liquid contains little empty space. Indeed, "…in a liquid a typical molecule is usually within the force fields of, say, 10 nearest-neighbor molecules and is never completely free of the influence of neighbors." (Berry, Rice, and Ross, 1980, p. 845).

The thermal velocity of matter at room temperature is some 6 Å per picosecond, more or less the speed of sound. Enormous thermal velocity in a system with little empty space implies a very high number of 'collisions' (using the word loosely) some $10^{13}$ times per sec if one dares to use a kinetic model of hard spheres to represent a liquid. The pores of channels are tiny, with diameters of (say) 4 Å and lengths of some 25 Å. Ions probably take some 100 nsec to cross that length, certainly they take more than 10 nsec (illustrated in the figures of Barcilon, Chen, Eisenberg, and Ratner, 1993).[9] Thermal motion leads to a great deal of thrashing about of a particle as it permeates the channel. It moves some 60 μm in 100 nsec, or (in other words) the particle moves randomly some 24,000 channel lengths before it crosses the channel. All this motion (60,000 nm total path length of random motion) produces little net displacement (2.5 nm net drift) because the ion reverses direction so many times (colliding) with neighboring atoms. On the microscopic[10] scale, friction and collisions dominate—as shown by elegant scaling arguments (Purcell, 1977; Berg, 1983)—producing a Brownian motion approximated by the random walk of a drunken man, the (limit of the) sum of frequent tiny displacements made by the ion with each atomic collision (Cox and Miller, 1965). The path of Brownian motion is a continuous function of unbounded variation, with an infinite number of maxima and minima, or zero crossings, in **any** finite interval, **no matter how small** the interval!

The random motion of atoms is very different from the smooth motion we see on the macroscopic time scale. It is even more different from the straight-line motion of the molecules of ideal gases (in which molecules hardly ever interact with each other and follow straight line ballistic paths—much larger than their diameter—between well defined collisions, collisions that themselves last negligible time).[11]

---

[9]See footnote 20.

[10]universally called "microscopic" although microscopes do not resolve atoms! I am told, however, by Tom DeCoursey that the usage fits with the Greek roots of the word "microscopic", viz. $\mu\iota\kappa\rho\acute{o}\varsigma$ = small and $\sigma\kappa o\pi\hat{e}\iota\nu$ = to view.

[11] For this reason, models of gas phase chemical reactions (e.g., Eyring rate theory) are not a good place to start a model of a condensed phase, at least in my view. Indeed, interactions are so pervasive in liquids that even the

Random motion of this sort is described by a Langevin equation (Gardiner, 1985) on the microscopic scale of individual atoms. On a macroscopic scale, it is described by the diffusion equation (Crank, 1975; Zauderer, 1983; Ghez, 1988; Kevorkian, 1990). The diffusive motion of ions through channels is at least as important as the drift motion in the electric field. Diffusive flux is driven by the concentration gradients of ions across channels; the corresponding description of individual random walks driven by concentration gradients is needed to link the (stochastic) motion of individual ions to (deterministic) macroscopic fluxes. As simple and important as this problem sounds, it has caused some difficulty because the relevant mathematical theory usually describes the motion of just one random walker, and does not in fact even use concentration as a variable! Only recently has concentration been included as a driving force in a stochastic theory of diffusion for the general situation of interest to biologists, namely for motion over a potential barrier of any shape (Eisenberg, Klosek, and Schuss, 1994, described later in this paper).

## *ELECTRODIFFUSION IN MIXED ELECTROLYTES*

The use of concentrations to approximate activities (i.e., free energy per mole) is nearly universal in biology and widespread in chemistry. It is often felt that the activity coefficient (the ratio of activity to concentration) is a fairly trivial correction, implying a different, albeit non-linear scale of concentrations but not having a qualitative effect on the equilibrium or nonequilibrium properties of channels (or solutions, for that matter). Ions are usually treated as independent entities with activities that depend on their own concentration, corrected by an activity coefficient that depends on total ionic strength of all species. The activity (in solutions of strong electrolytes, i.e., fully dissociated electrolytes) is nearly always assumed independent of the chemical nature of the other species in solution (i.e., independent of the concentration of a particular type of ion). This is a reasonable approach when dealing with solutions of one type of strong electrolyte, e.g., NaCl, but not when dealing with the mixed strong electrolyte solutions nearly always found in biology, as Duan Pin Chen has recently taught me. The usual assumption fails when

---

Boltzmann (transport) equation (Berry, Rice, and Ross, 1980: p. 1054) may not be a valid place to start analysis, because the idea of well-defined collisions is used in its formulation and derivation.

(1) ionic strength is varied.

(2) the composition of mixed solutions is varied (even at fixed ionic strength) as in "anomalous mole fraction" experiments (Hille, 1992).

(3) one constituent of a mixed solution far exceeds another, as is usually the case in biological solutions (e.g., NaCl $\cong$ 150 mM far exceeds KCl $\cong$ 5 mM in physiological solutions; $Ca^{++}$ is almost always much less concentrated than monovalent cations in physiological solutions. It is typically $2 \times 10^{-3}$ M outside cells and some $10^{-7}$ M inside cells).

Significant deviations from independence occur in all mixtures (even of 'ideal' gases: Mayer and Mayer, 1940, p. 203; Rowlinson and Swinton, 1982), not only electrolyte solutions, although there the effects are much larger. Mixtures of gases show strong departures from ideal behavior (e.g., Chapman and Cowling, 1970). Liquid mixtures cannot be understood "... solely from a knowledge of the  pure components. Such attempts rest upon the fallacy that the forces $(\alpha - \beta)$ between two molecules of species $\alpha$ and $\beta$ are always determinable from the strengths of the forces $(\alpha - \alpha)$ and $(\beta - \beta)$. If it were true that the $(\alpha - \beta)$ forces were always some 'average' of the $(\alpha - \alpha)$ and $(\beta - \beta)$ forces, then the properties of a binary mixture would be predictable in principle solely from a knowledge of those of the two pure components. However, such averaging is not universally valid. It is true that for very simple substances and for the prediction of relatively crude properties there are suitable averages .... However, such averaging is unsatisfactory for many classes of substances and inadequate for the detailed interpretation even of the simplest mixtures. One should rather take the observed properties of a binary mixture ...." (Rowlinson and Swinton, 1982, p. 86).

Ionic solutions show strong departures from ideal behavior in both their equilibrium and non-equilibrium properties. Mixed electrolyte solutions, such as the Ringer solutions that mimic the natural environment of cells have large excess free energy—a free energy of mixing beyond the ideal. In fact, a substantial fraction of the experimental work in classical electrochemistry was devoted to measuring and trying to understand this excess free energy and its variation with the species, concentration, and charge of ion.[12] Non-ideal behavior is even more apparent in the non-equilibrium properties of mixed electrolyte solutions, e.g., in their conductance. "It has long been known (Bray and Hunt, 1911), that the conductances of ***even very dilute solutions*** of mixed electrolytes are ***not additive***, so that Kohlrausch's law of independent ionic mobilities is ***not valid for mixtures***." (Kortüm, 1965, p. 211).[13] "Kohlrausch's rule of the independent migration of ions, valid as a limiting law for simple binary electrolytes, does not apply to mixtures, nor are the conductances additive." (Onsager and Fuoss, 1932).

Part of the non-ideality comes from a 'relaxation effect' arising from the ever present disturbance of the ionic atmosphere as an ion moves in an electric field. The resulting asymmetry in the ionic atmosphere means the charge density in front of the ion will be slightly lower, and that behind it slightly higher, than at equilibrium. Because ionic atmosphere has the opposite sign from the central ion, relaxation effects decrease the drift velocity.

Movement is also retarded by a 'electrophoretic effect' caused by the action of the electric field on the ionic atmosphere of a central ion. The ionic atmosphere itself is dragged in the opposite direction from the central ion, increasing their relative velocity, and so increasing the frictional retarding force on the ion. The properties of the ionic atmosphere depend on the species of ion and so both the relaxation effect and the electrophoretic effect are different for different species, even if they have the same charge.

These are just two of the many interaction effects in mixed ionic solutions. Others (appearing in higher order terms of the expansions of cluster theory: Anderson and Wood, 1973; Friedman, 1962; Mayer and Mayer, 1940) may be at least as important and yet not have been named or recognized as such. Indeed, present day theories do not fit the dependence of excess free energy on concentration, charge, or ionic species, very well, so clearly something is missing!

---

[12]I found the review of Anderson and Wood, 1973, (887 references!) and book of Tyrrell and Harris, 1984, a great help in dealing with the vast literature (cf., Franks, 1973). Substantial parts of Harned and Owen, 1958, and Robinson and Stokes, 1959, are devoted to mixed ion effects and their early analysis. Friedman, 1962, and Conway, Bockris, and Yeager, 1983, discuss more modern theories.

[13]Emphasis added. Note that 'very dilute' means less than 1 mM (!), Kortüm, 1965, p. 189.

From our point of view, the essential fact is that the activity and diffusion coefficient of one electrolyte is strongly affected by the concentration of another, even at fixed ionic strength. If the relative concentrations of ions were fixed, the effect of one ion on the activity of another might remain hidden from experimental view because it would be lumped into an 'activity coefficient' assumed (incorrectly) to depend only on the ionic strength, i.e., assumed (incorrectly) to be independent of the types of ions present in the mixed solution. But relative concentrations are not fixed in most biological experiments. Rather, they are varied and the effects of such variation are usually interpreted as properties of channels, not of interactions in the bulk solution.

The fact that the excess free energy, and conductivity, of mixed ionic solutions vary significantly with concentration and the resulting changes in driving force and conductance (of the bulk phase) certainly complicate traditional interpretations of physiological experiments. Indeed, some of the so-called anomalous mole fraction effect in channels (Hille, 1992) could arise from the excess free energy in the bulk solution, and not in channels at all (Chen, Ratner, and Eisenberg, personal communication).

The excess free energy $\Delta_m G$ of a solute depends on many variables. It depends on the amount of solute $A$, its mole fraction $y_A$ in a weight of solvent $W$ (in kilograms); and it depends on the second solute $B$, present in mole fraction $y_B$. The dependence is given by an empirical formula (Anderson and Wood, 1973), also derivable from cluster theory (Friedman, 1962).

$$\frac{\Delta_m G}{W} = RT \cdot I_s \cdot y_A \, y_B \big[ g_0 + (y_A - y_B) g_1 \big] \qquad (1)$$

where fixed ionic strength is $I_s$, $g_0$ and $g_1$ are constants, characteristic of the type of ion and solution, independent of the concentrations (i.e., mole fractions) of solutes $y_A$ or $y_B$ but dependent on ionic strength. $RT$ is the thermal energy per mole, with $R$ the gas constant and $T$ the absolute temperature. In simple solutions, like mixtures of KCl and NaCl, the $g_1$ term is negligible and the ionic strength dependence of it and $g_0$ is not significant.

The significance of equation (1) is clear; ***some of the driving force for the diffusion of one species is provided by the concentration of another*** and that excess driving force depends on the concentration of both species. The free energy of one species is coupled to the free energy of the other. ***Coupling is a phenomena of bulk solutions***, not just channels. In particular, if one solute in a solution is present in considerable excess (like NaCl in normal Ringer) or is a great deal more mobile than another, quite unexpected phenomena occur: "… [coupling] effects … are spectacular when the component is a strong acid …. The presence of the acid causes the salt to move against its own concentration gradient." (Tyrrell & Harris, 1984, p. 377).

The full implications of this excess free energy of bulk solutions of mixed electrolytes for the interpretation of channel experiments is not yet known; Duan Pin Chen uncovered this literature and its relevance to coupling only recently.[14] Even if the effects are relatively small in bulk solution, they may be large at the entrance of channels, where they influence offset potentials (i.e., the Donnan or built-in potentials, see Chen and Eisenberg, 1993) and concentrations, or the effects may be large within the channel itself. After all, the physical processes producing excess free energy in bulk solution are likely to occur in channels and at their entrances as well, albeit in different forms, because in channels the ionic atmosphere cannot translate spatially (Nitzan and Ratner, 1994), as it does in solution, although it certainly can deform.

Having identified the driving forces for ionic movement, present in any model of permeation, we turn to models of the permeation process itself.

## *MODELS OF PERMEATION*

Models of permeation are used whenever experiments are planned, data taken, or results analyzed. Whether the model is an explicit set of differential equations and boundary conditions, a computer program, a verbal analysis of states, or a vague image, the model exists in the mind of the experimenter, motivating (and thus governing) his choice of experiments and data to collect, and his analysis.

This role of theory is so embedded in the physical sciences that it often goes unstated. Physicists and astronomers since the time of Galileo and Newton have designed their experiments and chosen their observations with a theory in mind, usually an explicit theory at that. Chemists in the middle of the last century classified the atoms into a periodic table, a theoretical structure, and then designed experiments to test that structure. Since about the same time, chemists interested in molecules have designed their experiments to study their component atoms and their bonding. Indeed, 'molecules' are themselves theoretical constructs of atoms and the bonds between them. One of triumphs of science was the discovery that the periodic table was an output of the physics of the atom (quantum mechanics) and that chemical bonds arose from the same theory. One cannot imagine experimental chemistry today without the idea of a chemical bond. What once was a controversial theory has become the foundation of an experimental science, used (often unconsciously) by every scientist, even those exclusively interested in the cut and thrust of laboratory work.

---

[14]January, 1994. Interactions and coupling effects at the ends of channels have been previously postulated (Chen and Eisenberg, 1993) without a specific physical basis.

Indeed, the role of scientific theory is just as pervasive in some parts of our daily life. The computers and communication equipment we all depend on are the most striking examples, perhaps, because they are so clearly the reasonably direct result of rather abstract mathematics. But another example is less well known, although much older, and more supportive of our comfort and existence. The buildings we live in are designed on paper (and nowadays on computer screen) using simple formulae to deal with gravity, and (in the utilities of the building), electricity, water, and air flow. No one thinks of making a model of a building; the theory suffices. Indeed, it would be difficult to build a building if a realistic model were needed first. The model would cost nearly as much as the building itself and yet might differ from the real structure in significant ways. Of course, the theory of the building is not perfect. We all know of buildings where important functions have been left out (regrettably a common error of the more artistic and thus theoretical architects); and occasionally even the laws of gravity are ignored (for awhile, e.g., until the next earthquake). But all-in-all, construction is an applied science which has fully integrated its theoretical foundation.

Biology is not like that. For a variety of reasons, both historical and logical, many biologists pride themselves on the absence of biological theory, even those who deal with ideas much more than with laboratory results. Why biologists have so separated themselves from a main scientific tradition is an interesting question to be pursued in a historical or social context. Suffice it to say here, that theory flourishes in science after experiments are specific, quantitative and reproducible; after the underlying structures are known and understood, after major questions are agreed on. Before then, knowledge is vague and theories should be vague too. Vague ideas are usually worse than explicit theories, but sometimes it is the other way around. Vague ideas are better than specific theories when little is known, when much must simply be described. Vague ideas are better than specific theories if the theories are wrong, or (even worse) if theories are tautological, misleading, irrelevant or incomprehensible.

But enough is now known about the open channel that vague theories are unnecessary, indeed they are an impediment, at least in my view: experiments are specific, quantitative and reproducible; underlying structures are almost known although not understood; and the major questions are agreed on. The stage is almost set for a useful theory, that will embed itself as deeply into experimental practice as theory is embedded in the experiments of physics, chemistry, and engineering.

### State Models of Permeation are Inconsistent with the Electric Field

Until recently, channology was appropriately more devoted to the acquisition of data than its analysis. Indeed, such is still the case in most of molecular biology where the simple identification of the proteins of life is of the greatest importance and will certainly occupy most biologists for many years to come.

If the same approach is used to analyze the mechanism of proteins, however, it will not succeed. Analysis of mechanism requires an explicit physical model, one built on ***known*** properties and interactions of atoms.

The most popular models of permeation—indeed of any property of proteins—are state models (e.g., Hill, 1977, 1985; Hille, 1992). In these, an experimental observation is identified with a state of a protein, transitions from state to state are described by rate theory adapted from the Eyring theory of gas phase chemical reactions, and predictions are made and compared to experiments.

The difficulty with this class of theories is both fundamental and operational. Fundamentally, the difficulty is that the states and rate constants of these theories are vaguely defined, and they are rarely related to the fundamental physical processes that determine the properties of condensed phases. The functional dependence of states is usually not specified. How a state or rate 'constant' depends on the concentration, charge (permanent and mobile); on the trans-membrane potential and potential distribution; on the shape, size, structure, dielectric properties, flexibility (etc.) of a protein are the essential issues. Ignoring them is not helpful, I believe, if the goal is physical or atomic understanding of biological function. yet few of these variables appear in state models. Indeed, state models never (to my knowledge) compute the electric field. They assume (tacitly but all the more frighteningly for that) that fluxes can vary as concentrations and trans-membrane potential change without change in the shape of the electric field. I believe such models are fundamentally flawed; they are inconsistent with themselves and the first-order properties of the electric field.

> In this paper a 'consistent theory' is one that predicts an electric field consistent with the set
> of concentrations and charges in the theory, under all experimental conditions. More
> specifically, a theory is 'consistent' if, under any set of experimental conditions, the
> concentrations of ions in the system (i.e. channel, boundaries, and solution), when substituted
> into Poisson's equation (continuum models) or Coulomb's law (discrete models) predict the
> electric field **actually used** in that theory under those conditions.

State theories are hard to develop, to extend, to make more precise. Their vagueness means they can be easily fit to a range of data, but they cannot be easily falsified or tested. Their use tends to confuse description with understanding. They tend to form a closed universe quite disparate from the physical world known and described by chemists and physicists working in condensed phases. Indeed, it is often not clear how a state would be defined if everything were known, if the trajectory of every atom in a protein and pore were known.

Without a physical basis, rate constant models are both too flexible and too rigid. They are too flexible because additional states and rate constants can be added to fit awkward data. They are too rigid because the states and rate constants are treated as constants even if physics says they are variables. They are inconsistent with the first-order properties of the electric field. These are not desirable characteristics in a theory, let alone a paradigm motivating (and thus governing) most experimental work. In my view, a Kuhnian (scientific) revolution (Kuhn, 1970) is needed to replace the vague ideas of state models with the well defined, falsifiable, and thus experimentally useful ideas of the stochastic theory of electrodiffusion.

None of this implies that proteins do not have states, nor that their flux may not follow exponential laws in some cases. But those states should arise from a theory, with known functional dependence, or—much better yet—the states should be identified experimentally by ***direct measurement.***[15] Otherwise, it will be impossible to guess the states, or the laws which govern the rates by which the states change. And eventually investigators will be frustrated by the non-uniqueness of their theories and move on to other problems.

A theory in biology (or most other reductionist sciences) should have a vocabulary and syntax provided by another underlying science that has studied related (usually simpler, better defined, and better understood) systems in more detail. In that way, the vocabulary, syntax, and underlying ideas will have survived theoretical scrutiny ***and experimental test*** in the underlying field. Thus, physics draws its rules from mathematics (although it can choose which type of mathematics, and occasionally invent some, e.g., the Dirac delta function). Chemistry uses the laws of the electric field determined by physics. And molecular and atomic biology use the rules for molecular and atomic interactions in condensed phases. In particular, the grammar of molecular and atomic biology must explicitly include the electric field because that is the dominant force in most molecular and atomic interactions. Indeed, the electric field determines ***all*** the interactions of atoms and molecules, (according to the Hellmann-Feynmann theorem—Deb, 1981—as unlikely as that sounds) if the electric field is computed from the distribution of charge (i.e., electron density) that solves the Schrödinger equation (Bader, 1990; Szasz, 1992) and thereby minimizes the appropriate energy functional (Parr and Yang, 1989; Hohenberg and Kohn, 1964).

---

[15]The open and closed conductances of single channels are good examples of states well defined by direct measurement. The multiple closed states used to fit the kinetics of $Na^+$ conductance in nerve (the delay after a step in depolarizing—i.e., positive going—potential) are examples of unobserved, less well defined states.

In the case of proteins, the appropriate grammar is physical chemistry of the condensed phase (Hirschfelder and Curtis, 1954; Hansen and McDonald, 1986; Gray and Gubbins, 1984; Nitzan, 1988); electrochemistry (Kortüm, 1965; Bockris & Reddy, 1970; Israelachvili, 1985); and the dynamics of proteins (Brooks, Karplus, and Pettitt, 1988; McCammon and Harvey, 1987). That grammar uses physical variables as nouns, variables like potential, location, charge, concentration, and flux. It uses as verbs the differential equations of condensed phase physics, basically a combination of Newton's laws and Maxwell's equations with a smattering of quantum mechanics added. Indeed, the verbs are simply the laws of electrodiffusion (on one length scale or another) if we consider only the open channel, as we do here.

The need to consult an underlying science, to use its vocabulary and syntax, is not just a moral rubric. The lack of knowledge of an underlying field can have direct effects on every day experimentation. Physiologists like me have spent a good fraction of their careers changing the mixture of components in electrolyte solutions (blissfully) unawares of the work of electrochemists on such mixtures.[16] Indeed, biologists in general, including physicians, and some chemists, are taught to treat ionic solutions ideally, as mixtures of independent components, although ionic solutions of biological interest do not behave ideally in the laboratories of electrochemists. Not knowing of these non-ideal properties of bulk solutions, we could not evaluate their significance to biological phenomena. We do not know what part of the effects of solution changes used routinely in channology occur in the bath! Knowledge of the underlying science is needed to find out.

In a similar spirit, the theory of condensed phases can sometimes provide justification for state theories. Such state descriptions are extremely helpful if they are correct, if they can be derived from the grammar of physical theories just described with explicit functional dependence on the parameters of the molecules, etc. Such is sometimes possible, even for quite general situations (see Eisenberg, Klosek, and Schuss, 1994, discussed below). But such derivations cannot be generally expected, nor can they ever be guessed with any certainty. ***States should be the mathematically derived output, not the assumed input of analysis***.

---

[16]See footnote 12.

If states are used as inputs to a theory, it is easy to assume properties that proteins cannot have. For example, if states are assigned *a priori* without physical basis, it is natural to make the corresponding rate constants independent of the concentration of permeating ions. But, as we have seen, such independence would imply a potential of mean force independent of substrate concentration and that is impossible under any range of conditions. The substrate shields the permanent charge of the protein; the shielding is a function of substrate concentration and the potential of mean force and the corresponding rate constant will be a function of substrate concentration.

## MAKING MODELS IN ATOMIC BIOLOGY

Evidently kinetic models with rates and states are not to my taste. What then are? How do we go about making a consistent model of a channel? What are the objects of the model, its nouns? What are the dynamics of the model, its verbs? Those are the questions we now turn to.

Ever since atoms were imagined, and temperature was understood to be their random motion (Brush, 1976), physicists have struggled to relate macroscopic observables to microscopic quantities. The difficulties are formidable in the gas phase, for spherical atoms (Chapman and Cowling, 1970; Hirschfelder and Curtis, 1954; McCourt, Beenakker, Köhler, and Kuščer, 1990); in the liquid phase, they are much worse (Hansen and McDonald, 1986); for a solution of charged particles in a highly polar and asymmetrical solvent (Gray and Gubbins, 1984; Martynov, 1992), the problems are insurmountable in my opinion: the system deviates from ideality in too many different ways; each deviation is very hard to deal with individually—in fact, few of the deviations can be approximated to more than a leading term, yet higher order terms may be expected to (collectively) dominate, as they do for multipole expansions of the electric field. And, as the *coup de grâce,* each non-ideality interacts with the others in ways that are likely to be large, yet very hard to formulate, analyze, or solve *a priori*.[17]

Understanding these problems, many despaired of a (general, accurate, reasonably simple) theory of ionic solutions. But such a theory may not be necessary. Computers may allow us to reconstruct the actual atomic motions of a solution, and thus simulate the properties of the solution, without a general theory. That is the goal of molecular dynamics.

---

[17] Indeed, how can we even be sure that all significant interactions have been identified and included?

## *MOLECULAR DYNAMICS*

Molecular dynamics simulates a solution by describing each of its atoms and their interactions.[18] The nouns of the simulation are the atoms and chemical bonds of the molecules. The verbs are Newton's laws of motion. The forces between atoms (either nouns or verbs depending on your choice) are described by a force-field, given by the derivative of so-called potentials (not to be confused with electrical or thermodynamic potentials) which summarize the properties of chemical bonds (Scheraga, 1968; Burkert and Allinger, 1982; Rigby, Smith, Wakeham, and Maitland, 1986; Maitland, Rigby, Smith, and Wakeham, 1987; McCammon and Harvey, 1987; Brooks, Karplus, and Pettitt, 1988; Dykstra, 1993). These potentials can be determined reasonably well when the distribution of charge in each molecule is quite constant, when 'polarization' of an atom or molecule (corresponding to what we have called induced charge earlier in this paper; in this case, the induced charge is within an atom or molecule and is caused by movement of its cloud of electron orbitals) is small and the range of temperatures and phenomena are limited (e.g., to exclude quantum effects like changes in covalent bonds). In that case, molecular dynamics is able to predict the properties of pure liquids quite well (Davidson, 1993; Haile, 1992; Allen and Tildesley, 1989). If the molecules are asymmetrical, difficulties arise, because long range effects begin to occur: raisins are found at the top of boxes of cereal because shaking of a random homogeneous mixture of inhomogeneous objects does ***not*** necessarily produce a homogeneous result. If long range electric fields are present, these probably need to be separately incorporated into a simulation[19] as we shall see. If solvent and solute are present, asymmetrical, and charged, the calculations become difficult to manage, because of the needed size of the system and duration of the integrations. But workers are hopeful that the exponential development of computers will permit the simulation of solutions, and thus the solution of solutions, in the near, if not immediate future.

---

[18]The November, 1993, issue (Volume **93** (7)) of ***Chemical Reviews*** (e.g., Davidson, 1993) contains a most informative series of review articles on molecular dynamics of many different systems.

[19]Probably by linking to another level in a hierarchy of models (see below).

Extending these techniques to proteins (Brooks, Karplus, and Pettitt, 1988; McCammon and Harvey, 1987) is irresistible, inevitable, and valuable, even though approximate, so approximate in the view of some that it has attracted serious criticism. In my view, the critics probably have a point, but they should stimulate, not prevent work: the molecular dynamics of proteins provides a more realistic and detailed view of atomic motions than many of us ever thought possible, and that view will be of enormous value in understanding the function of proteins, as well as in constraining lower resolution models to decently approximate atomic reality.

The question for channels is then obvious: why not just calculate the motion of an ion as it permeates a channel? Indeed, why not just simulate a system with one concentration of ions on one side of the membrane, another concentration on the other, with an electrical potential as well? Why not calculate enough trajectories to determine the flux and be done with the problem once and for all?

The answers are all the same "I wish we could, but we can't."

One of the reasons we can't is that oversimplified 'potentials' are used to describe atom-atom interactions. The choice of potentials determines the force-field around atoms and so significantly modifies the electrostatic barriers for permeation. It also determines the difficulty of the calculation and so simplifications are usually made, which neglect induced charge, and which lump hydrogens together with a more massive neighbor into a composite pseudo-atom, whose force-field implicitly includes the effects of the hydrogens. However, *ab initio* quantum mechanical analysis using density functional theory (D.P. Chen, C. Lee, and H. Chen, personal communication) show that polarization of hydrogens by the electric field dramatically reduces energy barriers for permeation. Computations of gramicidin that include all hydrogens ***explicitly*** (e.g., the hydrogen in the -$CH_2$ groups of the protein) predict an order of magnitude smaller barrier than traditional calculations in which hydrogens are not explicitly present and do not polarize (for the most part). Kumpf and Dougherty, 1993, have shown that "cation-$\pi$" interactions between permeating ion and aromatic residues of channel proteins are large and specific, suggesting that they may be involved in ionic selectivity. It is obvious that quantum mechanical calculations of channels, being only recently conceived, is just reaching infancy. Rapid growth and a productive childhood are expected, along with a stormy, hopefully brief adolescence, and a maturity that cannot yet be imagined.

The other problems of molecular dynamics discussed here arise directly from limitations in the size and duration of the calculation. Molecular dynamics is extremely useful—indeed, indispensable for the understanding of short range interactions, like those that probably dominate ion entry into channels, selectivity, and perhaps flux coupling. Molecular dynamics is also needed to determine the meaning, limitations, and parameters of other more approximate treatments. But we shall see that molecular dynamics *per se* is restricted, like any other technique. I spend disproportionate time highlighting these limitations trying to curb over-enthusiasm, trying to focus molecular dynamics on the questions it can answer, using other techniques to address other questions. But highlighting limitations is also a step towards understanding and removing them: I hope some reader will be stimulated to remove limitations that seem unavoidable to me!

***Temporal Limitations***

It seems likely that an ion takes some 10 to 100 nsec to permeate a channel[20], that is to say, to reach the opposite side for the first time (see figures in Barcilon, Chen, Eisenberg, and Ratner, 1993). A direct simulation of permeation would then need to be at least this long to observe one elementary event. In practice, it might have to be much longer, if ion movement in the bath must be included, if the approach and retreat of the ion from either side of the channel is also of interest.

Simulation of permeation requires computation of an ensemble of permeation events under each condition of interest. An ensemble might mean 2,500 events if an accuracy of $1/\sqrt{2500} = 2\%$ were acceptable. A typical current-voltage 'curve' would contain at least one

---

[20]The current through many open channels is a few picoamps. One picoamp corresponds to one elementary charge every 160 nsec. Many workers think that open channels contain just one ion at a time.

hundred such points and so would require simulations of 12.5 msec altogether, calculated as shown below:

$$50 \frac{\text{nsec simulation}}{\text{trajectory}} \times 2{,}500 \frac{\text{trajectories}}{\text{predicted points}} \times 100 \text{ points} = 12.5 \text{ msec}$$

A present-day high speed workstation calculates about 10 psec of simulation per hour of computer time, for a channel like gramicidin with some 200 atoms. One current-voltage curve would thus take about a billion hours. Computing would have to speed up about a billion times to make this calculation practical.

Of course, prolonged calculations are not just limited by computer time; they are often irreproducible as well because of the discrete nature of computer arithmetic. Round-off error is a significant limitation (Haile, 1992; Allen and Tildesley, 1989). In our (Chen and Eisenberg) experience, individual trajectories of the channel protein gramicidin longer than (say) 10 psec are likely to be sensitive to details of machine arithmetic. Long calculations also show behavior reminiscent of chaotic systems (Hoover, 1991; Allen and Tildesley, 1989). Individual trajectories are exponentially sensitive to the assumed initial positions of atoms after some (say) 10 psec. The initial positions are of course not known with much accuracy, and the initial velocities are more or less completely unknown: even a fine x-ray structure might have an RMS error in position of 1 Å, and x-ray measurements barely constrain (and do not determine) velocities at all. These uncertainties are enough to ensure striking divergence in individual trajectories after some psec of dynamics.

Individual trajectories should not be the focus of attention. Rather, it is the ensemble of trajectories that determines the prediction of experimental events and data. The assumption widely made (e.g., Hoover, 1991) is that averaging over an ensemble of trajectories removes the errors of discrete arithmetic and the sensitivity to initial position, etc. This assumption is inevitable and underlies any large scale calculation of mechanical systems (like molecular dynamics). And it is certainly reasonable and likely to be true in many cases. Nonetheless, it cannot always be true.

Systems with different qualitative behavior in different regions of phase space, exquisitely sensitive to initial conditions, are called 'chaotic' nowadays and are common in large scale nonlinear mechanical (Bai-Lin, 1984; Moon, 1987) and chemical (Scott, 1991; Brumer, 1988) systems, some quite like channels (Hoover, 1991). The differential equations of molecular dynamics form such a system, exceedingly non-linear by the standards of mechanical systems. Sometimes chaotic systems have periodic behavior, yet have trajectories that seem a random function of time.[21] Sometimes one region of phase space cannot be accessed from another at all, or not in finite time (Allen and Tildesley, 1989). It is not clear how a numerical simulation would average over such systems. It is not clear what the average value of a quantity would mean, and estimators of variance and other sample statistics (like power spectra, correlation functions, or coherence functions) would be even more problematic. Of course, such a system would have bizarre behavior in the experimental world as well as in the world of simulation. Thermodynamic experiments, *performed in finite time,* might well not produce estimates of the thermodynamic parameters of such a system. It might take infinite time either to compute or observe those properties. So it is not clear how to study this kind of system, given the time available to most of us. What is clear is that ***theory, simulations and experiments must describe the same state space*** and that state space should be sampled in the same way in each.

To summarize, perhaps in too vivid words: the computational ensemble (of round-off error) is not always the thermodynamic ensemble (of everything); and neither may coincide with the ensemble measured in experiments.

### *Spatial Limitations*

The size of the system computed is also a serious limitation on molecular dynamics (Haile, 1992; Allen and Tildesley, 1989). Calculations today are limited to something like 10,000 atoms. And that certainly seems, at first blush, to be enough to calculate the short range interactions in a system. However, systems of interest are three-dimensional and so even a 27,000 atom system would have just 30 atoms in each direction. Obviously, at second blush, one would wish a bigger system. Just as obviously, exponential growth in computer capability will give us much bigger systems very soon.

---

[21]e.g., commercially available (pseudo)random noise generators (made from internally fedback shift registers) are a practical example, Horowitz and Hill, 1989. They produce an apparently random signal, with a (nearly) flat power spectrum, yet the signal is *strictly* periodic.

> ***Periodic boundary conditions*** are often used to increase the spatial size of simulations of equilibrium (but not non-equilibrium) systems: Allen and Tildesley, 1989; Hoover, 1991; Haile, 1992. Periodic boundary conditions assume that all the independent variables of the calculation (e.g., the positions of the atoms *and their velocities)* are the same on each corresponding face of an array of space filling solids. That is to say, the system is decomposed into cubes and the independent variables are assumed to have the same value at a particular location on all corresponding faces of the cubes.

> Periodic boundary conditions cannot be applied to the nonequilibrium case—at least as usually implemented—because they do not permit a gradient of concentration, electrical potential, or electrochemical potential between corresponding boundaries.

> Periodic boundary conditions are dangerous even in the equilibrium case—at least in my view—for two reasons.

> - First, ***they replace actual physical boundary conditions*** with artificial periodic conditions. The physical boundary conditions often have large, even dominant effects on the properties of the system (as we see repeatedly in this paper) and so replacing them is likely to change the system qualitatively.

> - Second, ***periodic boundary conditions force the system to be artificially rigid***. They ensure that the system is (spatially) periodic in the velocity and position of every one of its atoms at every time. That is to say more precisely, ***all the co-ordinates of phase space, at all times*** are exactly the same (at corresponding positions on corresponding boundaries), thus imposing an impressive rigidity, a rigidity that real matter does not have, not even a crystal at a temperature of absolute zero, in which only the average positions of the atoms are periodic.

What is not often realized, is just how big the system needs to be to simulate a system like ionic channels involving concentrations and potentials. It is clear that the system needed to simulate channels has to be big enough to define an average concentration and average electrical potential. That is unfortunately very big. Certainly, one would wish 2,500 solute atoms in the system; that means some 600,000 water atoms on both sides of the channel if the solute is $Na^+$ in typical Ringer solution, much more if the solute is $K^+$ or $Ca^{++}$. That is large, frighteningly large, but perhaps computable.

The volume needed to define a concentration can be frighteningly large, but the volume needed to define an electric field can be still much larger. The electric field needs to be computed in the volume of space containing relevant charge—*whatever volume that is*. A cubic volume some 20 Debye lengths on a side might seem sufficient under biological conditions, say a volume $20 \times 3$ Å $= 60$ Å on a side, containing somewhat less than the number of atoms necessary to define a concentration. That may be a reasonable estimate of 'the relevant volume' in free solution. But *that estimate ignores the effect of boundaries and the charge along them*. The volume enclosed by those boundaries can be very large indeed ($cm^3$) in cases where the system is surrounded by an insulator, like a cell membrane. And neither that charge nor the potential it creates at the boundary is always fixed; rather they vary and interact nonlinearly with (potentially) all the other charges in the system. Simplifications separating boundary from system may be possible, but are not obvious and need analysis for justification.

Consider for example, a real biological application, the simulation of a single $Na^+$ channel in a nerve cell, a nerve membrane wrapped into an indefinitely long cylinder, a cable some 10 µm in diameter if it is a vertebrate nerve, or some 500 µm diameter if it is a squid nerve. The opening and closing of that channel, and the current through the channel once it is open, are found experimentally to depend on the potential across the membrane, if that is controlled by the standard electronics of a patch clamp (on one side of the membrane: Sakmann and Neher, 1983) and a voltage clamp (of one sort or another, Hille, 1992; Smith, Lecar, Redman, and Gage, 1985) on the other. But if the nerve is left in its natural state, without the voltage clamp, then the current through the single channel, and the gating of the channel, will depend on the potential within (roughly speaking) one length constant of the channel.[22] In the mammalian nerve, the length constant would be millimeters; in the squid nerve, it would be centimeters. A direct simulation of either case is out of the question: it would involve perhaps $10^{21}$ ions, and 100 times more particles, for the simulation of the squid! This explosion in size occurs because a nerve fiber is designed by evolution to act as a transmission line, an imperfect capacitor, which stores charge along a macroscopic length $\lambda$, with little decrement in potential. In other words, the charge that determines the potential in a particular channel molecule is spread over a millimeter to centimeter distance of the nerve fiber.

---

[22]The length constant $\lambda$ is the parameter used to describe the spread of potential in cylindrical cells. It is a measure of the effectiveness of the membrane as an insulator.

Transmission lines like nerve fibers have been studied for a long time—even before Maxwell wrote his equations of the electric field—ever since Kelvin (1855, 1856) analyzed the spread of charge and potential over thousands of miles to help design the transatlantic (telegraph) cable (Gray, 1908; Smith and Wise, 1989). And similar equations[23] (without inductance, however) have been used by physiologists to model the (electrically) linear properties of nerve and muscle fibers for nearly that long. Hodgkin and Huxley, 1952, showed how to use the differential equations of cable theory to describe the long distance active conduction of charge along a nerve fiber (the nerve impulse or action potential—nearly a soliton—that propagates and carries information) while using a localized description of (active nonlinear) current flow through the nerve membrane, or through its channels, as we would say today. They realized that two different theories were needed with two different resolutions. One theory (e.g., their theory of ionic conductances[24]) described the channels, another described the cable of the nerve. The output of the high resolution theory (of ionic conductances) was a trans-membrane (i.e., radial) current that served as the input of the low resolution theory of conduction itself.

I feel that the same approach, using models with multiple resolutions, is needed to construct an atomic theory of ionic channels and connect it to biological reality. It seems obvious that the flux through a single channel will not depend on the microscopic properties of all the ions needed to simulate its electric field. Surely, the flux will depend in some average way on the ions far from the channel. The problem is how to make such a low resolution theory of the charges far from the channel, how to link it to the model of the channel with its atomic resolution, how to be sure the models are mutually consistent.

---

[23]The same equations describe the transmission lines of electrical engineering (Ghausi and Kelly 1968; King, 1965) and nerve and muscle fibers of animals (Jack, Noble, and Tsien, 1975) and are called the telegrapher's equations in applied mathematics (e.g., Zauderer, 1983). Note these equations are used with inductance set to zero in biology, in contrast to the situation of interest to Kelvin and his successors.

[24]The October, 1992, special (supplementary) issue of ***Physiological Reviews*** "Forty Years of Membrane Current in Nerve" is a most useful description of progress in this field (e.g., Gardner, 1992).

### *HIERARCHY OF MODELS OF THE OPEN CHANNEL*

An hierarchy of models is needed—each on a different scale, each consistent with the electric field on that scale—to cope with the immense range of sizes and times between a channel (angstroms) and a nerve fiber (centimeters), between the vibrations of a protein's bonds (femtoseconds) and the average time course of gating (milliseconds to seconds). ***Only with a hierarchy of models can the atomic structure of a channel be linked to its macroscopic functions.***[25]

Molecular dynamics is indispensable for the analysis of short range forces and phenomena. And these are of the greatest importance in understanding channels, including their critical properties of selectivity, flux coupling, and ionic interactions. But molecular dynamics is only one of a hierarchy of models needed to deal with experimental data, to link structure and biological function. Molecular dynamics is the foundation of the analysis of channels and proteins and that foundation needs to be firm if the super-structure—the house of the atomic biology of channels— is to stand tall; everything depends on the foundation, but channels (and people, if I may continue to torture the metaphor) live and function in the house, the structure itself, not in the basement, and so we ought not neglect the higher levels. Each level of the hierarchy rests on its foundations, namely the lower levels. The parameters of each higher level—an analysis with less resolution— will be the results of a lower level.

---

[25]The need for a hierarchy of models has no doubt been apparent to many for some time (e.g., Jakobsson, 1993). I proposed such a hierarchy—unknowing of any other work along these lines, if such did in fact exist—in a grant application "A Hierarchy of Models of the Open Channel" submitted to the NSF in the summer of 1992 and then organized a workshop that presented the ideas at the Biophysical Society the next winter (Eisenberg, 1993).

In particular, the outputs of a simulation of molecular dynamics will be the inputs of a theory of stochastic dynamics, as we shall now see.[26]

### *Stochastic Motion of the Channel: Langevin Dynamics*

Our next landing on the staircase of the hierarchy of models is stochastic dynamics, or Langevin dynamics. Here, the dynamics of a permeating ion are described by

1) the Langevin equation (Gardiner, 1985; Arnold, 1974; Schuss, 1980; Gard, 1988; Doob, 1953; Karlin and Taylor, 1975), an equation giving the position $\mathsf{X}(t)$ of the particle at time $t$ or

2) the intimately related Fokker-Planck equation giving the probability density function of finding a particle near position $x$ with velocity $\dot{x}$. The Langevin equation is a form of Newton's second law (that the sum of the forces equals the [acceleration]×[mass]) with an added random term. The forces in the Langevin equation are (a) the (deterministic) electric field that produces the drift of charged particles; (b) the (deterministic) retarding force of friction; (c) the random term describing the fluctuations in force produced by the microscopic collisions that generate the deterministic friction.

**Langevin Equation**

$$-m\frac{d^2\mathsf{X}}{dt^2} = m\beta(x)\frac{d\mathsf{X}}{dt} + ze\Phi'(x) - \sqrt{2mkT\beta(x)}\dot{w} \qquad (2)$$

I write this expression in the customary terse form found in the literature with the amusing mixture of notations we have inherited for derivatives (Boyer, 1949: e.g., p. 252; Cajori, 1980, e.g., p. 257; Grabiner, 1981), writing some derivatives with Leibniz' differential quotient notation $\frac{d\mathsf{X}}{dt}$, others with Newton's fluxion notation (superscript dot) $\dot{w} \equiv \frac{dw}{dt}$, and still others with

---

[26]The channologist less interested in our struggles in the theoretical hierarchy may wish to turn ahead to the section *Flux Ratios.* The more general reader may wish to turn even further to the section *IS THERE A THEORY?*

Lagrange's prime notation $\Phi'(x) \equiv \dfrac{d\Phi(x)}{dx}$. The effective mass of the moving particle is $m$; $m\beta(x)\dfrac{d\mathsf{X}}{dt}$ is the friction term, with $\beta(x)$ the friction coefficient (per unit mass), $ze\Phi'(x)$ is the electrical force, with $z$ the valence of the particle (a positive or negative number), $e$ the charge on a proton ($1.6 \times 10^{-19}$ cou), and $\Phi(x)$ is the potential of mean force (a functional also of many other variables besides the location $x$). $\sqrt{2mkT\beta(x)}\dot{w}$ is the random force, where $kT$ is the thermal energy, typically some $4 \times 10^{-21}$ joules $\cong 0.96$ kcal $\cong 0.025$ $e$·volts, where $k$ is the Boltzmann constant and $T$ the absolute temperature, and $\dot{w}$ is the derivative of standard Brownian motion as defined in, e.g., Doob, 1953; Billingsley, 1986; Karlin and Taylor, 1975; Gardiner, 1985.

Condensed phases like biological solutions are dominated by collisions and friction because they contain little empty space to accommodate the large thermal velocity of their atoms and molecules. An enormous number of atomic collisions occur in the shortest time of interest and so it is natural to consider large friction expansions of the relevant dynamical equations. The usual high friction expansion of the full Langevin equation drops the second derivative term from equation (2) (Hänggi, Talkner, and Borkovec, 1990). The resulting reduced Langevin equation is the starting point for analysis of stochastic diffusion phenomena and is thought to describe the trajectories that account for the macroscopic diffusion described by Fick's law. It is a great deal easier to handle both numerically and analytically than the full Langevin equation (2).

**Reduced Langevin Equation**

$$m\beta(x)\frac{d\mathsf{X}(t)}{dt} = -ze\Phi'(x) + \sqrt{2mkT\beta(x)}\dot{w} \qquad (3)$$

The equivalent high friction expansion of the Fokker-Planck equation yields the Smoluchowski equation (of the same form as the Nernst-Planck equations given later in this paper) and predicts a Maxwellian distribution of velocities.

The variables of the Langevin equations (2) & (3) are written here in the conventional, if misleading manner. The variables are not as independent as the notation, naively applied, implies. In particular, the electric field $-ze\Phi'(x)$ is a functional (e.g., a solution of Poisson's equation with boundary conditions) that depends on nearly everything else in the problem, because several of the types of charge described previously in this paper depend on nearly everything in the problem. The structure of the protein determines the permanent charge that helps determine $\Phi(x)$: note it does not determine the potential directly since a protein is a distribution of charge not potential. The protein is not connected to a source of energy and charge like a battery so it cannot maintain a potential in the face of changing charge, whether that is caused by changes in experimental conditions, boundary conditions, or properties of the protein itself.

Dynamics of the protein itself would enter chiefly in better approximations than equation (3), e.g., the dynamics of the protein would determine the memory kernel (describing the 'impulse response of the friction function) in a generalized Langevin equation (Hynes, 1986; Gardiner, 1985). The friction and mass terms may also have complex dependencies, although these are more obscure since their description is not as clear as the electric field. The electric field is produced by charge, and only charge (in the present problem), but the effective mass and friction reflect a multitude of interactions and dynamics in the system, evident in simulations although not well understood in analytical theories of liquids and solutions.

### *Simulations of the Reaction Path: the Permion*

The Langevin equation determines a stochastic path that describes one ion flowing in a channel, like a particle in a chemical reaction. The ensemble of those random paths is the distribution described by the Fokker-Planck equations. The lowest energy path of that distribution is called a reaction path and can be computed by methods developed (for proteins) by Ron Elber (Czerminski and Elber, 1990; Nowak, Czerminski, and Elber, 1991) that estimate the path of steepest descent in the potential surface describing a chemical reaction.

The reaction path through a channel is often assumed to be a straight line, but calculations (Elber, Chen, Rojewska, and Eisenberg, 1994) of the reaction path within gramicidin show rich behavior. The path found in the calculations[27] is quite different from a straight line, with deviations of the order of 1 Å, enough to have energetic effects of many *kT*. The simulations show that the straight-line Cartesian co-ordinate $X(t)$ of the Langevin equations (2) & (3) needs to be replaced with a curvilinear reaction path.

---

[27]excluding the entrance and exit regions which we have not yet simulated.

Along the reaction path, there is significant motion of all of the waters within the channel's pore and of most of the atoms of the channel protein, as the permeating ion moves through the channel. It is useful to consider all these motions together, defining a quasi-particle—a **permion**— much like the quasi-particles holes, phonons, and electrons (for that matter) defined in solid state physics (Kittel, 1976; Sze, 1981; Shur, 1990; Seeger, 1991) to represent correlated motions of many atoms of a semiconductor.

Many properties of the permion can be evaluated along its reaction path. The effective mass (as customarily normalized and defined, e.g., Czerminski and Elber, 1990) is much less than the mass of the corresponding permeating ion. The friction is quite complex, strongly dependent on location within the channel, and with interesting time dependence as well. Finally, the calculated free energy is an estimate of short-range interactions in this region of the channel. More realistic descriptions of the free energy are problematic since it is not clear how to simulate the entry process or the different types of electric charge described previously in this paper. Direct evaluations of the noise term—its frequency spectrum and position dependence—would also be interesting.

Calculations of dielectric effects are needed. Perhaps some insight could be developed by calculating the free energy of (hypothetical) ions of different (fractional) charge, but otherwise similar properties. Such ions can be created in the simulations of molecular dynamics by progressively 'turning off' the charge on the atom, leaving all its other parameters the same. The dependence of free energy on charge is the physical property that defines dielectrics and allows the definition of a dielectric constant, when the dependence is linear. Analysis of that dependence in complex non-ideal systems like proteins may allow generalization of the idea of dielectric 'constant' (Rogers, 1990). An operational definition of dielectric constant for complex systems might be a useful description of the energy stored in the distortion of its electric field, that is, in its induced charge. We can seek such a definition by studying the variation of free energy as the charge of an atom is changed in a simulation of molecular dynamics, all other properties of the atom being kept the same. If the free energy is a linear function of the assumed charge, and robust—i.e., reasonably insensitive to the details of the calculation—the slope of the function should allow definition of an effective dielectric constant. If the slope is not linear, or is overly sensitive to details, such a definition is not likely to be very useful; indeed, it might be misleading.

In this way, molecular dynamics can supply paths, parameters and functions (like the potentials, frictions, and masses that depend on position) for use in a Langevin equation. Of course, molecular dynamics also provides physical insight (e.g., the nature of interactions, the existence of a curvilinear path) and an atomic perspective on protein motions (the nature of dehydration and resolution on entry; the existence of a permion) that could not be acquired from other techniques. But *molecular dynamics needs to be supplemented with solutions of the Langevin equation if macroscopic variables like concentration and membrane potential are to be introduced* into a theory of the open channel, so we can try to describe experimental phenomena (like current flow) that so clearly depend on those driving forces.

### *Langevin Dynamics*

Langevin dynamics (Arnold, 1974; Gardiner, 1985; Schuss, 1980; Gard, 1988)—often in the high friction limit, equation (3)—is the standard starting place for most stochastic theories of chemical reactions (Nitzan, 1988; Hynes, 1986; Steinfeld, Francisco, and Hase, 1989). It has the considerable advantage of simplicity, compared to molecular dynamics—it appears to be an analytical theory, although the stochastic nature of $w$, and its bewildering properties (Billingsley, 1986; Doob, 1953) (it is a continuous function of unbounded variation, with an infinite number of maxima and minima, or zero crossings, in any finite interval, no matter how small the interval!) makes exploitation of the analytical theory difficult: for example, boundary conditions are difficult to prescribe or interpret for a function of unbounded variation, even if it is continuous.

The direct application of the reduced Langevin equation (3) to channel problems has been impeded by a simple if frustrating problem: current through channels is driven by gradients of concentration (as we have often said), but concentration does not appear in either Langevin equation, (2) or (3); rather they describe the trajectories of the particles one at a time and the corresponding Fokker-Planck equations describe the probability that *one particle* is found in some region at some time. This problem is particularly frustrating in view of the central role of concentration in diffusion and the enormous (Hänggi, Talkner, and Borkovec, 1990) literature of stochastic motion over potential barriers. So many difficult issues have been confronted and solved in that literature that it seems as if the problem of introducing concentration into a stochastic model of diffusion should have been solved long ago.

The problem has, however, plagued simulations of stochastic dynamics of channels, ever since they were started by Kim Cooper working in Erik Jakobsson's lab (e.g., Cooper, Jakobsson and Wolynes, 1985; Jakobsson and Chiu, 1987, 1988; Cooper, Gates, and Eisenberg, 1988a,b; Chiu and Jakobsson, 1989; Cooper, Gates, Rae, and Eisenberg, 1989; Gates, Cooper, and Eisenberg, 1990; Barcilon, Chen, Eisenberg, and Ratner, 1993). It is not clear how to introduce particles into the simulation in a way that mimics ion entry from a solution of fixed concentration, or how to remove particles from the simulation in a way that mimics ion exit from a channel, into a solution of fixed concentration.

When we first looked at this problem, we found that most papers in the literature of stochastic dynamics assumed large barriers for diffusion (Hänggi, Talkner, and Borkovec, 1990; Steinfeld, Francisco, and Hase, 1989; Nitzan, 1988; Hynes, 1986; Gardiner, 1985; Risken, 1984) to reduce the problem to a study of diffusion at *just* the top of a barrier. Rates over the top of the barrier were then scaled to the appropriate concentrations in a way we did not fully understand but no doubt can be justified (e.g., see Barcilon, Chen, Eisenberg, and Ratner, 1993: Section VIII, eq. 8.6). Our analysis could not use that approximation because flux through channels does not depend just on properties at the top of potential barriers within the channel. It is known experimentally to also depend strongly on the potential (i.e., charge) and concentrations far away from the channel interior, for example, at the boundaries of the system, as we have discussed at length.

Furthermore, I believe on evolutionary grounds that barriers in channels are probably not large, and certainly should not uncritically be assumed large early in the development of a theory[28]: many channels are designed to conduct as much current as possible for their biological function, and those are the ones we can observe with presently available patch clamp amplifiers. It seems likely that the channels we study today will have small barriers, not large ones, particularly those involved in the nerve action potential.[29] We were also reluctant to follow this approach because of our previous experience with other boundary value problems. In those, boundary conditions proved to be of the greatest importance, often determining the main properties of the system actually observed experimentally. In the present problem, of stochastic motion over a potential barrier, we knew if the usual assumptions were made about the potential, the direct effect of boundary conditions disappeared, submerged beneath the high barrier (Barcilon, Chen, Eisenberg, and Ratner, 1993: Section VIII, eq. 8.6). We feared the consequential loss of physical meaning.

It seemed clear that we would have to derive our own theory, starting with the existing literature of the stochastic motion of a single particle and extending it to concentrations of particles moving over barriers of any shape. Surely, we thought, it would be easy to generalize the theory of one particle to a theory of $N$ particles, at least if they were assumed not to interact. All we thought necessary was to multiply the probabilities of the usual theory by $N$, the number of particles.

The trouble was the more we understood of the problem, the more were the possible choices of 'the probability' and the 'number of particles'. Which probability? The probability of which ion, where, conditional on what? Which number of particles? The number of which ion, on which side of the channel, conditional on what? We could not guess. Indeed, as we learned still more, we found other ways to compute flux in the literature, for example, by determining mean first passage times (MFPT) and channel contents (Naeh, Klosek, Matkowsky, and Schuss, 1990). We determined these times and contents using some identities and analytic tricks (Barcilon, Chen, Eisenberg, and Ratner, 1993) but found apparent paradoxes. The MFPT could be defined in different ways, each with different mathematically imposed boundary conditions (none of the boundary conditions were determined by the physics of the problem) and each definition of MFPT gave conflicting expressions for the fluxes, some nonsensical. Probability theory is well known to be subtle and difficult (Szekely, 1986; Romano and Siegel, 1986; Stigler, 1986) but it certainly should give unique results in a well defined physical set-up like this, which is the experimental system originally analyzed by Fick, 150 years ago!

---

[28]I am afraid that once an assumption is made in a theory, and becomes firmly established in the experimental literature, it becomes very hard to evaluate critically or to remove the errors it causes, particularly if direct experimental tests of the assumption are not possible.

[29]Evolution has gone to considerable trouble to maximize the conduction velocity of the propagating nerve impulse by (for example) growing giant nerve fibers in invertebrates or wrapping nerve fibers in insulating myelin in vertebrates. It seems likely that evolution would go to at least as much trouble to maximize the flow of $Na^+$ ions (which is the energy and current source for propagation), if it possibly could.

The problem was particularly frustrating because the history of diffusion theory suggested the solution should be straightforward. After all, the original problem posed by Fick involved diffusion from one concentration to another (actually through a biological system, it is amusing to note in the present context[30]). The description of the stochastic trajectories underlying this classic diffusion should be straightforward, we thought, if we simply followed the classical work of probabilists, starting with Einstein, on diffusion and Brownian motion.

For some six years, Victor Barcilon and I struggled with this problem, quite unsuccessfully. Finally, we found a way to avoid it, to solve part of it without understanding the stochastic basis of the solution, by computing the average flux using an analytical trick (the construction of a formal adjoint) which was clearly true, depending only on the accuracy of our algebra, but which was empty of stochastic meaning. We—joined now by Mark Ratner and Duan Pin Chen—compared the analysis with simulations with satisfactory results and then Duan Pin Chen showed how to simplify the simulations, deriving statistical formulae of tantalizing simplicity, but whose probabilistic meaning we could only guess, not derive (Barcilon, Chen, Eisenberg, and Ratner, 1993).

Understanding required a stochastic theory, later constructed with Zeev Schuss and Malgorzata Klosek (Eisenberg, Klosek, and Schuss, 1994). We tried for nearly two years just to formulate the problem in stochastic language, seeking to describe a concentration boundary condition at the ends of the channel by invoking the special properties of that location (e.g., Chen and Eisenberg, 1993a,b) in channel proteins. The problem was that trajectories do two different things at a boundary with a defined concentration. Some trajectories start there and others end (i.e., are absorbed) there. The problem was how to describe these behaviors in the language of stochastic trajectories. Only a few boundary behaviors of stochastic trajectories have been analyzed or are thought to be possible (Karlin and Taylor, 1975; Lerche, 1986; Gut, 1988). The problem was to map ***both*** the starting and ending behavior of trajectories (at one location of known concentration) into those boundary behaviors. We tried for a long time to separate starting and ending trajectories by invoking special properties of the channel protein, i.e. barriers or 'selectivity filters' at the ends of the channel's pore. But we did not succeed because (as it turned out) we were always working entirely in the high friction limit.

---

[30]Fick was a physiologist interested in membranes, as well as diffusion (Hille, 1992; Jacobs, 1935).

The key to solving the problem was realizing that it had nothing to do with channels! In fact, traditional stochastic theories of diffusion could not describe diffusion from a region of one concentration to a region of another, whatever the link between them, whether it was a channel or free solution. In other words, the classical stochastic theory of diffusion did not solve the original problem of diffusion posed by Fick.

The difficulty arose from the classical (and our previous) description of diffusion ***in the high friction limit***. We (along with many others studying Brownian motion in condensed phases since Einstein) had started with the reduced Langevin equation (3) as a description of diffusion. We had assumed that diffusion in a solution is so entirely dominated by friction[31] that the second derivative could be dropped from the full Langevin equation (2) leaving the reduced equation (3), describing trajectories with a distribution of position, but not velocity. The corresponding Fokker-Planck equation (really, the Smoluchowski equation) showed that the distribution of velocity was in its steady state equilibrium configuration, namely Maxwellian.

---

[31]because there is little empty space in a solution yet particles move at thermal velocities of some 6 Å per picosecond ensuring collisions occurring every 100 femtoseconds, while observations are made only every microsecond, at the very fastest.

If we described the particle by a reduced Langevin equation, with only one spatial derivative, we were entitled to only one boundary condition no matter how complicated were the processes at the ends of the channel. Yet at a boundary of fixed concentration, trajectories have two behaviors: trajectories both start and end there. The place of known concentration behaves both as a source and a sink. These behaviors are quite different and cannot be described by a single boundary condition. We needed an equation that allowed two boundary conditions. Then, we could solve or simulate the problem for a wide range of behaviors at the channel mouth, indeed, we could simulate it for any behavior that could be defined in the formalism of the Langevin equations.

That equation was easy to provide, once we realized it was needed. The full Langevin equation itself provides the extra boundary condition. It involves both the velocity and the location of a particle and describes trajectories of particles with a distribution of both. We had to use the full Langevin equation and deal with it, even when friction dominates movement of ions, as in classical diffusion theory. Evidently, the distribution of velocities is ***not*** Maxwellian, even in the high friction limit, even to first approximation, in the non-equilibrium case of interest to us, when flux flows through channels with low barriers.[32]

At first, the prospect of dealing with the full Langevin equation seemed daunting, at least to me. The full Langevin equation, or its fraternal twin the full Fokker-Planck equation, is not noted for its mathematical simplicity. Analytical solutions are rarely available, and even the general properties of the solutions are not known (Gardiner, 1985; Risken, 1984; Hänggi, Talkner, and Borkovec, 1990). Imagine my pleasure then when we found identities, valid for potential barriers of any shape or size, describing the flux from one concentration to another, e.g., through a channel.

$$J_{net} = J_{in} - J_{out} = \frac{1}{\sqrt{2\pi\varepsilon}} \left( C_L \cdot \text{Prob}\{R|L\} \ - \ C_R \cdot \text{Prob}\{L|R\} \right) \tag{4}$$

---

[32]The analytical solution to the problem demonstrates this point explicitly. The distribution of velocities deviates from the Maxwellian ***whenever*** flux flows, even when friction is overwhelmingly large.

$\varepsilon$ is dimensionless temperature, named to be consistent with the literature of the Langevin equation (see precise definitions in Eisenberg, Klosek, and Schuss, 1994). It is not a small number despite its name! $C_R$ and $C_L$ are the concentrations on the right and left, respectively. The probabilities $\mathsf{Prob}\{R|L\}$ and $\mathsf{Prob}\{L|R\}$ are the solutions of precisely defined stochastic problems (described in Eisenberg, Klosek, and Schuss, 1994) that can solved analytically in some simple cases and simulated in all others (Barcilon, Chen, Eisenberg, and Ratner, 1993).

This formula is closely related to the traditional flux formula of the Nernst-Planck equation.

$$J_{net} = J_{in} - J_{out} = \frac{C_L e^{\Phi(0)/\varepsilon}}{\int_0^1 \frac{e^{\Phi(\zeta)/\varepsilon}}{D(\zeta)} d\zeta} \quad - \quad \frac{C_R e^{\Phi(1)/\varepsilon}}{\int_0^1 \frac{e^{\Phi(\zeta)/\varepsilon}}{D(\zeta)} d\zeta} \tag{5}$$

But now the integrals could be derived in terms of conditional probabilities of specific well-defined stochastic problems. The probabilities $\mathsf{Prob}\{R|L\}$ and $\mathsf{Prob}\{L|R\}$ of equation (4) are just the integrals of equation (5), if the dynamics are dominated by friction. If the dynamics are more complex, the conditional probabilities can be evaluated by simulations of Langevin equations or molecular dynamics.[33]

The analysis of the stochastic problem does more than provide a stochastic interpretation of the well known Nernst-Planck flux formula. It also reveals the role of *cis* trajectories, namely *LL* and *RR* previously unknown (at least to me) perhaps because they do not contribute to the mean steady-state fluxes of isotope customarily measured in experiments. *LL* trajectories do not contribute current (if the patch electrode is on just the outside, which we call the right). But they have a large effect on channel properties because they occupy the channel a substantial fraction of the time and so block other trajectories (in channels with interactions). *RR* trajectories block the channel in a similar way. They also contribute directly and substantially to the variance (i.e., open channel noise) even though they contribute zero mean current. When an ion flows from the electrode into the channel it produces a short blip of current in one direction; when it flows back into the electrode, completing its *RR* trajectory, it contributes a blip in the other direction. The *RR* trajectories significantly complicate analysis because they (unlike the others) do not form a stochastic process that can directly analyzed by renewal theory (Cox, 1962). The *cis* trajectories *RR* and *LL* are likely to be important under normal conditions; indeed, they may dominate the normal open channel noise in at least some channels (Hainsworth, Levis, and Eisenberg, 1994). If a 'slow' ion is present (i.e., an ion with a small diffusion constant) the *RR* and *LL* trajectories are expected to determine most of the behavior observed experimentally.

---

[33]Dynamics can be complex for many reasons, e.g., (1) if the ions and channel are described by molecular dynamics; (2) if ions interact with each other; (3) if they significantly modify the shape of the electric field; or (4) if friction or potential is 'frequency dependent' so a generalized Langevin equation is needed to describe ionic motion.

The stochastic analysis also is closely related to traditional results on average fluxes:

- The flux law (5) is identical (in that special case of mean fluxes in the steady-state) to the integrated Nernst-Planck equation, the flux law assumed for radioactive isotopes by physiologists ever since the time of Ussing and Hodgkin (1949, or thereabouts).

- The probabilistic analysis yields the familiar formula (Naeh, Klosek, Matkowsky, and Schuss, 1990).

$$\text{Flux} = \frac{\text{Content of System}}{\text{Mean First Passage Time}} \qquad (6)$$

provided the flux, mean first-passage time MFPT, and contents—here of the channel—are the appropriate conditional quantities defined in Eisenberg, Klosek, and Schuss, 1994.

- The flux law is identical (for the mean fluxes in the steady-state) to the flux law derived previously with the adjoint method by Barcilon, Chen, Eisenberg, and Ratner, 1993 and their simple statistical formulae turn out just to be the 'best' estimators of the conditional probabilities $\text{Prob}\{R|L\}$ and $\text{Prob}\{L|R\}$ of the flux law (4)!

## *CHEMICAL REACTIONS*

Our stochastic model of the open channel has a general significance that caught me by surprise: it permits permeation (through the open channel) to be described ***exactly*** as a chemical reaction, moving ions by rate constants between well defined states, namely definite concentrations at specific locations.

$$L \xrightleftharpoons[k_b]{k_f} R \tag{7}$$

Such reactions have been written qualitatively, or in terms of vaguely defined states $L$ and $R$ for many years. But now we can define the rate constants exactly, for barriers of any shape and height, and the meaning of $L$ and $R$ is clear and unambiguous.

$$k_f \equiv \mathrm{Prob}\{R|L\}; \qquad k_b \equiv \mathrm{Prob}\{L|R\} \tag{8}$$

with the probabilities defined by the integrals in equation (5) above if dynamics are simple, or, if they are complex[34], by simulations of the underlying stochastic problems specified in Eisenberg, Klosek, and Schuss, 1994.

This probabilistic interpretation of a chemical reaction with prescribed concentrations at its boundaries is not confined to channels and channel permeation. Rather, it applies to any chemical reaction in which the concentrations $L$ and $R$ are maintained fixed, for example, it applies to any chemical reaction in a continuous flow system (Naumann and Buffham, 1983), even if the reaction occurs over low potential barriers.

We were surprised to find a probabilistic model of a chemical reaction exactly ***valid for any shape potential barrier***. Most theories of a chemical reaction assume a particular shape barrier, in particular a high barrier (Hänggi, Talkner, and Borkovec, 1990; Steinfeld, Francisco, and Hase, 1989; Nitzan, 1988; Hynes, 1986; Gardiner, 1985; Risken, 1984) and I thought that was a necessary assumption. It is certainly a reasonable assumption for the typical chemical situation involving distinct chemical species; after all, in a certain sense it is the high barrier that defines the species and makes them distinct.

---

[34]See footnote 33.

The high barrier also has a profound effect on the role of boundary conditions in the problem. The high barrier ensures that flux is determined (asymptotically) by diffusion *just* over the top of the barrier. Boundary conditions do not occur at that location, however, and so (in a certain too vivid sense) the usual analysis hides boundary conditions behind its high barriers.

In the analysis of channels, barriers cannot be assumed large, as we have seen[35] and so boundary conditions emerge into the prominence they usually have in physical problems. Inclusion of boundary conditions often complicates a problem, and adds terms to its solution, and so we were surprised to find such simple expressions (8) & (5) for the rate constants. These arise because of the boundary conditions we use. If the same chemical reaction[36] took place in a different experimental set-up (with concentrations not held constant and so with different boundary conditions), the reaction would *not* follow exponential kinetics (in time) and so could not be described by simple forms of the law of mass action.

## WHAT WAS WRONG?

It is natural to wonder what is different about the present analysis from other previous analyses (including my own) of the high friction case. The difficulty in earlier analyses was in the way friction increased in the limiting process. The limit of high friction was taken while keeping the overall driving force fixed (i.e., keeping fixed the difference in electrochemical potential, or electrical potential, or concentration, depending on the precise problem). Thus, the flux necessarily went to zero as the friction increased, implying that the system approached equilibrium in that high friction limit. The limit desired and used in the present analysis (i.e., in Eisenberg, Klosek, and Schuss, 1994) is quite different. We study high friction, but with a fixed amount of flux (independent of the friction), a flux corresponding to that observed in experiments.

Subtleties often occur in limiting processes when multiple parameters are present: indeed, singular perturbation theory (Kevorkian and Cole, 1981) was designed to deal with such situations. In general, a limiting process ought to correspond to the precise experimental situation of interest. Otherwise, the result of the limiting process is unlikely to describe the experimental situation, even qualitatively. In the present case, the limit of high friction and fixed driving force is inappropriate because it implies an equilibrium (or tiny perturbation from equilibrium) quite different from the situation in real open channels. In real channels, the limit of high friction occurs in a system far from equilibrium in which flux (of at least one ion) is nearly always substantial. Limiting a diffusion system too close to equilibrium is particularly dangerous according to Hoover, 1991 (p. 266[37]): "The change from equilibrium to non-equilibrium is … not a simple one that can be described by [regular, as opposed to singular] perturbation theory. The phase-space geometry is fundamentally changed in a way which frustrates traditional analyses based on Taylor series expansions …. Away from equilibrium *only* rare [fractal] states can correspond to nonequilibrium steady-states" (italics are in original).

---

[35]See footnote 29.

[36]i.e., the chemical reaction (7), with the same reactants, with the same dynamics over the same potential barrier, with the same diffusion constant and other parameters.

## BACK TO THE HIERARCHY

In this way, we have learned how to include diffusion in a stochastic model, how to predict the effects of concentration gradients. This step on the staircase of the hierarchy is secure. But it is an isolated landing, standing by itself because the stochastic analysis assumed a potential function $\Phi(x)$; it did not compute it. This model is not consistent[38] as we have defined it. The analysis is only valid under a range of conditions in which the potential function does not change, i.e., under a range of conditions that do not change the various types of charge described previously. Such conditions exist and can be studied experimentally, but they are quite restrictive, and certainly do not span the domain of biological interest.

Clearly we need to compute the potential in a consistent stochastic model from the physical description of the charges and the Langevin description of dynamics. That is an essential step in the construction of the hierarchy needed to analyze the atomic physics of channels, linking atomic structure to biological (macroscopic) function. But we do not know how to do that calculation (although, of course we have ideas, or should I say hopes?). For now, we can compute the electric field and the concentration coupled together only when we give up the discrete description of ions. To determine coupling, we have, at least up to now, described the electric field as a fluid, specified by the Poisson equation, and the concentration of each type of ion also as fluids, specified by the Nernst-Planck equations, all with appropriate boundary conditions. With that departure from reality, the entire problem can be analyzed, with charges producing potentials, potentials producing charges, both producing concentrations, and all producing flux. That approach has proven most fruitful in the analysis of semiconductor devices (Jerome, 1994) and we will apply it now to channels. It is the highest, but most remote level of the hierarchy dealt with here.

---

[37]Hoover considers stochastic diffusion systems quite similar to those studied here.

[38]Precisely defined earlier in the text, see page 29.

To summarize, in our over-worked metaphor: the continuum model is the attic of our house of hierarchy, giving a good view of an attractive horizon, but too musty and remote from our foundations (of molecular dynamics) to inhabit any longer than we have to!

## *MACROSCOPIC MODELS OF THE CONTINUUM*

Macroscopic models are rather unfashionable in molecular biology, particularly in the study of single ionic channels, where they seem faintly archaic and out of place, given that measurements are made every day of current flow through single channel molecules. Nonetheless, the resolution of single channel measurements may make life seem more complicated than it is. General principles may be obscured by atomic detail, just as too much magnification makes a photograph too large and objects hard to recognize. A photograph showing each molecule of pavement will not show the road. The Navier-Stokes equations can be hard to see even in the Boltzmann transport equation let alone in the trajectories of molecular dynamics. A theory of an ensemble of channels might be simpler and more closely related to the underlying physics of channels than a stochastic theory of one channel!

Too little resolution is, of course, even worse than too much, because it can blur two different mechanisms into a hybrid very different from either. Consider a mechanism we all know first hand, reproduction. If human anatomy were studied with too little resolution, say by observers from another planet, males and females might be lumped together into one composite structure, a homogeneous blur of an inhomogeneous population. In that blur the mechanism of reproduction would be lost. Until the ***functionally*** relevant populations were distinguished, and separately studied, the mechanism of reproduction would be entirely misunderstood. Indeed, until this error were corrected, more experimental work would lead to more confusion, not knowledge! Until we know where in the hierarchy of models evolution has made its adaptations, we need to study all levels. We will not know which models are over-magnified and which are too blurred until we are finished.

### Levels of the hierarchy: Occam's razor can slit your throat

If a complex system is studied or analyzed on only one level of the hierarchy, the (most understandable, necessary, but illogical) human need to solve something—to feel we have at least some control over our uncontrollable world—can lead us to accept the 'simplest' explanation, using Occam's razor[39] to justify our analysis, even if it is irretrievably blurred by low resolution of essential structures or events. If a theory ignores or does not well resolve the length scale relevant for a particular function of the system, it will not describe it properly, even if the theory is simple and fits some set of experimental data. No theory of reproduction that ignores the distinct morphology of sexes can be correct, no matter how well it fits a set of experimental data. Or (to choose an example closer to hand), no theory of the action potential in nerve fibers that lumps together $Na^+$ and $K^+$ currents will resolve the properties of the individual channel molecules. ***Occam's razor slits your throat if theories and experiments do not have enough resolution***, if they blur together essentially separate components of the underlying mechanism, if they describe the wrong level of the hierarchy of models.

Macroscopic models seemed blurred and inappropriate descriptions of channels because ions, water molecules, and the channel's pore are about the same size. Ions and water moving through a channel clearly form a single file system (for example) and it seems ridiculous to represent or analyze such systems as a continuum. Nonetheless, macroscopic models describe some properties of channels quite well, particularly those that depend on long range phenomena like the electric field.

---

[39]"Plurality is not to be assumed without necessity" or "What can be done with fewer assumptions is done in vain with more" (as quoted in Pece, 1994).

Continuum models have an important role because ***they are consistent and can be completely analyzed***. Indeed, at the present time, they are the ***only*** models that can treat the electric field consistently. Only in continuum models have all the sources of the electric and concentration fields—that is to say, all the charges present and all the sources and sinks of matter—and all boundary conditions been included and their effects analyzed by mathematics, with minimal additional assumptions. Furthermore, interactions appear automatically from the mathematics—without *ad hoc* assumption or analysis—if a complete set of field equations and boundary conditions are specified. The different types of interactions do not have to be specified, or at least enumerated so explicitly, as in discrete models, and subtle and high-order interactions which are consequences of fundamental laws arise automatically from the mathematics and will be found in the output of the calculations, even if the interactions are not manifest in the differential equations and boundary conditions that define the problem, even if the existence of interactions was not anticipated by the authors of the model, even if the physical basis of the interactions is not understood *a priori*.

The importance of this mathematical power is great. Nonlinear coupled partial differential equations behave in complex ways often unanticipated even by people who have studied them for decades. Their behavior is often an unanticipated mystery. Later, after the complex behavior of a system is understood physically, it may seem obvious, natural, often inevitable. In nonlinear coupled systems, rarely is the 'obvious' or 'inevitable' predicted until after it is computed. The behavior may be simple and reproducible, and most useful technologically (e.g., a mechanical switch) but if it arises from ***essentially*** nonlinear coupling between underlying equations, it is nearly impossible to predict *a priori* by analysis, verbal or mathematical.[40]

---

[40]for most of us. John Bardeen was clearly an exception (Holonyak, 1992; Pines, 1992).

Thus, the reasons for making macroscopic consistent models of the open channel are

- The sources of the electric field are understood.

- The sources and sinks of concentration are completely understood.

- The boundary conditions are (fairly) evident and quite well understood.

- A wide range of interactions are predicted automatically by the mathematics, even if they are not explicit in the original formulation of the problem.

- A complete analysis of a wide range of experimental conditions can be made using only mathematics, without making additional physical or biological assumptions.

A complete consistent model can be made (solved and computed) of an hypothetical macroscopic system, an open channel filled with a continuous electric fluid, a fluid with charge density and concentration, not discrete ions.

We suspect that all effects found in such a continuum model are present in a real channel filled with discrete charges. Certainly the physics that is important in a continuum model should be present in a more realistic discrete reworking of the theory. If we find important effects of one type of charge in the continuum model, it should be included in the discrete model. If we find important effects of one type of boundary condition in a continuum model, it should be included in a discrete model. If we find important interactions in a continuum model, the physics producing them should be included in a discrete model.

In fact, we have found, rather to our surprise, that the interactions in a continuum model of an open channel can account for a wide range of complex behavior, behavior that previously had been thought impossible unless the system were discrete. The richness in behavior occurs because the electric field changes the concentrations and concentrations change the electric field. The electric field and concentration are coupled. Neither can be analyzed without the other.

Most experimental interventions change the shape of the electric field and thereby have a large effect on permeation. Most previous theories of the open channel assume the electric field is independent of experimental conditions, or varies in a simple pre-ordained way, and so miss these first-order effects, which can dominate even the qualitative behavior of the channel.

Changes in the shape of the electric field are used in semiconductor devices. They perform most of their functions that way. They amplify, switch ON and OFF (i.e., gate), and do logic as the potential (or current flow) at one boundary changes the electric field everywhere. *If the electric field were assumed of constant shape, transistors and integrated circuits could not be designed or understood*. Unable to formulate or solve a discrete model, physicists have for nearly fifty years designed semiconductor devices using a consistent macroscopic model (Shockley, 1950; Sze, 1981; Shur, 1990; Seeger, 1991; Mock, 1983; Markowich, 1986; Hess, 1988; Jerome, 1994) similar to a continuum model of open channels. They too would dearly love to formulate and solve a consistent discrete model; but the continuum model has been good enough to support the exponential growth in capability, speed, and complexity that makes much of our modern technology possible.

## *POISSON-NERNST-PLANCK (PNP) MODEL*

In a continuum model of an open channel, each ionic species is represented as a fluid, a concentration that is a function of time and location. The sources of the fluid are the electrodes far away at infinity. The fluids—the concentrations of ions—can flow under a variety of forces. In channels, the main fluxes are thought to be diffusion and drift in gradients of concentration and electrical potential. Convection is conventionally ignored[41] and the system is supposed to be uniformly isothermal on all length scales. A thorough analysis demonstrating the universal insignificance of convection and temperature would be comforting!

The equation governing flow is the Nernst-Planck equation which relates the flux $J_j$ (moles$\cdot$cm$^{-2}\cdot$sec$^{-1}$)—independent of location in this steady-state situation—of species $j$ with charge $z_j$ to the diffusion constant $D_j(x)$ (cm$^2$ sec$^{-1}$) at location $x$; with constant $RT$, the thermal energy per mole at room temperature.

---

[41]in the bath. In the channel, coupling between ion and water flux has been included in many phenomenological models.

$$J_j = -D_j(x)\left[\frac{dC_j}{dx} + \frac{z_j F}{RT}C_j\frac{d\Phi}{dx}\right] \qquad (9)$$

The driving forces of gradients of concentration $C_j$ and electric potential $\Phi$ are often written as shown without their list of arguments, but **this notation is seriously misleading**. Because the concentration(s) of ions help determine the potential (through Poisson's equations) and the potential helps determine the concentration (through the Nernst-Planck equations), both variables depend on the entire properties of the system everywhere, not just at one location. Thus, $\Phi_j\left(x; C_j[\tilde{x}]\right)$ might be an appropriate notation, showing that the potential depends on location $x$, but also on the electric field **at every other location** $\tilde{x}$, as well. Similarly, the concentration of a given species $C_j(x)$ may also depend on the concentration $C_k(\hat{x})$ of each other species $k \neq j$ at every location $\hat{x}$, and also on the potential $\Phi_j(\bar{x})$ at every location $\bar{x}$, as well. The appropriate notation is then something like $C_j\left(x; \Phi[\bar{x}]; C_{k \neq j}[\hat{x}]\right)$. Of course, both the concentration and the electric field depend on the diffusion constants, dimensions of the channel, and all other parameters in the system. While it is understandable that a simplified nomenclature is often used—and I use it myself as much as possible—it is regrettable that the underlying complexity of the physics is sometimes forgotten.

The Nernst-Planck equation can be written in terms of the electrochemical potential $\tilde{\mu}_j(x)$ and either the electrical potential or the concentration

$$\tilde{\mu}_j(x) \equiv RT \ln \tilde{C}_j(x) + z_j F \Phi(x) \qquad (10)$$

by neglecting the effect of one ion on the free energy of another, the mixed ion effects mentioned previously. Note that the potential is measured with respect to a reference point to be chosen and set to zero potential, the solution outside the channel on the right hand side, in the usual convention. Similarly, the (dimensionless) concentration $\widetilde{C}_j(x)$ is measured with respect to a reference concentration $\overline{C}$, a standard state, so

$$\widetilde{C}_j(x) \equiv \frac{C_j(x)}{\overline{C}} \tag{11}$$

Later we will measure the flux with respect to the same standard concentration, introducing the flux $\widetilde{J}_j$, measured in units of $\overline{C}$, defined by

$$\widetilde{J}_j \equiv \frac{J_j}{\overline{C}} \tag{12}$$

The Nernst-Planck equation can be written in terms of the electrochemical potential $\widetilde{\mu}_j(x)$, the electrical potential $\Phi(x)$, and the dimensionless flux $\widetilde{J}_j$.

$$\widetilde{J}_j = -D_j(x)\left[\exp\left\{-z_j F\Phi(x)/RT\right\}\right] \times \frac{d}{dx}\left[\exp\left\{\widetilde{\mu}_j(x)/RT\right\}\right] \qquad x_1 \le x \le x_2 \tag{13}$$

Remember that flux is independent of location and time in this steady-state model of an open channel and so this expression (13) can be explicitly integrated between any two locations $x_2$ and $x_1$ of its domain of validity.

$$\widetilde{J}_j \int_{x_1}^{x_2} \frac{\exp\left\{z_j F\Phi(x)/RT\right\}}{D_j(x)} dx = \exp\left\{\widetilde{\mu}_j(x_1)/RT\right\} \;-\; \exp\left\{\widetilde{\mu}_j(x_2)/RT\right\} \tag{14}$$

giving what we call the ***Integrated Nernst-Planck*** equation:

$$\underbrace{\widetilde{J}_j}_{\textbf{\textit{Net Flux}}} = \underbrace{\frac{\exp\left\{\widetilde{\mu}_j(x_1)/RT\right\}}{\displaystyle\int_{x_1}^{x_2} \frac{\exp\left\{z_j F\Phi_j(x;C_j[x])/RT\right\}}{D_j(x)} dx}}_{\textbf{\textit{Unidirectional Efflux}}} - \underbrace{\frac{\exp\left\{\widetilde{\mu}_j(x_2)/RT\right\}}{\displaystyle\int_{x_1}^{x_2} \frac{\exp\left\{z_j F\Phi_j(x;C_j[x])/RT\right\}}{D_j(x)} dx}}_{\textbf{\textit{Unidirectional Influx}}} \tag{15}$$

The numerators of the integrated Nernst-Planck equation (15) depend only on the electrochemical potential in the baths, where the concentrations and electrical potentials are known and under experimental or biological control. The numerator is entirely independent of the electric field and is **un**coupled, despite the nonlinear coupled nature of the overall system.[42]

The denominator of the integrated Nernst-Planck equation (15) depends explicitly only on the electric field and is independent of the concentration of a tracer ion in the bath (present in negligible concentrations—when measured in moles/liter, not disintegrations per second). For that reason, the ratio of the terms in equation (15), namely the flux ratio, is particularly useful, as we discuss later in this paper. The flux ratio (originally introduced by Ussing, 1949; see Hodgkin and Keynes, 1955a,b; Schultz, 1980; Hille, 1989; Sten-Knudsen, 1978; Jacquez, 1985) should be independent of the shape of the electric field within the channel. Indeed, the ratio of fluxes should be given by the (exponential of) the difference in electrochemical potential on the two sides of the channel, quite independent of what goes on in the channel.

$$\frac{Unidirectional\ Efflux}{Unidirectional\ Influx} = \exp\left\{ \left[ \tilde{\mu}_j(0) - \tilde{\mu}_j(1) \right] \Big/ RT \right\} \equiv \textbf{\textit{Ideal Flux Ratio}} \qquad (16)$$

Here, $\tilde{\mu}_j(0) - \tilde{\mu}_j(1)$ is the difference in electrochemical potential between the inside and outside bath, the overall driving force for electrodiffusion determined by the difference in electrical potential and the difference in (the logarithm of) concentrations (really free energies per mole, namely activities).

The unidirectional fluxes of the main species can easily be estimated from measurements of the total fluxes of radioactive tracers (indeed, that is why Ussing defined them!). Each term can be experimentally identified by measurements of the total flux of radioactive isotope, when isotope is present only on one side of the channel (in trace amounts nonetheless easily detected by its radioactivity), and is (essentially) absent from the other (see Jacquez, 1985; Chen and Eisenberg, 1993b for details). The flux ratio determined from equation (15) gives such powerful results—a ratio quite independent of what goes on inside the channel or of the shape of the electric field in the channel—because the trace amount of radioactive isotope used to estimate macroscopic unidirectional fluxes does not change the denominator of the equation (15).

---

[42]It would be interesting to try to recover the integrated flux equation from a discrete model, particularly to see how the various interactions and couplings disappear as the macroscopic model arises from its more realistic discrete cousin.

The meaning of unidirectional fluxes and their ratio became clearer to me when I tried (unsuccessfully) to define them in another way, when I realized that the Nernst-Planck equation can be written in terms in terms of electrochemical potential and concentration as well as in the traditional form shown in equation (13). Thus,

$$J_j = -\frac{1}{RT} D_j(x) C_j(x) \frac{d\widetilde{\mu}_j}{dx} \quad ; \qquad x_1 \leq x \leq x_2 \qquad (17)$$

This expression can be integrated between any two locations $x_1$ and $x_2$ in its domain of validity.

$$J_j = \frac{\widetilde{\mu}_j(x_1)}{RT \int_{x_1}^{x_2} \frac{1}{D_j(x)C_j(x)} dx} - \frac{\widetilde{\mu}_j(x_2)}{RT \int_{x_1}^{x_2} \frac{1}{D_j(x)C_j(x)} dx} \qquad (18)$$

The numerators of this expression (18) depend only on the electrochemical potential in the baths, and it was instructive to try to define each term as a unidirectional flux, and the ratio of terms as a unidirectional flux ratio, as we have done previously with equation (15). Such a definition is quite different from the traditional one and gives quite different results. The apparent paradox is easily resolved by applying classical (e.g., Jacquez, 1985; given once more in Chen and Eisenberg, 1993b) analysis of unidirectional and tracer flux to equation (18). Then, we see that holding the concentration of tracer (nearly) zero on the *trans* side of the channel (which is the key step in relating the net flux of tracer to the unidirectional flux of the main species) does not set either of the terms in equation (18) to zero. Furthermore, the denominator of equation (18) is different for each species of ion[43] and so the denominator does not cancel if the flux of one species (e.g. tracer isotope) is divided by the flux of another (e.g., the main species). For these reasons the flux of isotope does ***not*** estimate one of the terms in equation (18), nor is the ratio of terms in equation (18) robust and revealing.

---

[43] strikingly different, if one of the species is held at zero concentration on one side, as the tracer species usually is.

Returning now to the classical analysis—the flux ratio (16)—we notice that it, the ratio, is uncoupled from the electric field, that is to say, independent of the potential profile $\Phi(x)$ within the channel, although the fluxes themselves are not. The terms in the traditional integrated Nernst-Planck equation (15) are coupled because the denominators of that equation depend on the potential $\Phi(x)$ everywhere within the channel, and thus on all the properties of the channel, but the ratio of terms does not.

The potential within the channel—the function $\Phi(x)$ in the denominator of (15)—is generally unknown, except as the solution of Poisson's equation, which in turn depends on the Nernst-Planck equation. The denominator may be the same for every ion, but it is the same unknown value that depends on the potential everywhere $\Phi(x)$, which also depends on the concentration $C_j(x)$ of every ion everywhere: $\Phi(x)$ depends on almost every parameter and variable in the system. For that reason, the ***integrated Nernst-Planck equation*** (15) ***cannot be used until the electric field is specified***.

The electric field is specified by ***Poisson's equation***.

$$\frac{d^2\Phi(z)}{dz^2} = \underbrace{-2\widetilde{\omega}_0(z)}_{\substack{\text{Permanent}\\\text{Charge}}} - \underbrace{\widetilde{q}(z)}_{\substack{\text{Channel}\\\text{Contents}}} - \underbrace{2\widetilde{\varepsilon}\Delta(1-z)}_{\substack{\text{Constant}\\\text{Field}}} - 2\widetilde{\varepsilon}\Phi(z) \qquad \overset{\substack{\text{Induced Charge:}\\\text{Dielectric Effect}}}{} \tag{19}$$

where $\Delta$ is the trans-membrane potential and $\widetilde{q}(z)$ is the sum of the concentration of all ions at location $z$. The channel extends from $z = 0$ to $z = 1$. See Chen and Eisenberg, 1993a,b for precise definitions.

This one-dimensional version of the Poisson equation was derived from the full three-dimensional system and boundary conditions (Barcilon, 1992; Barcilon, Chen, and Eisenberg, 1992; Chen, Barcilon, and Eisenberg, 1992) where the system of dimensionless units is described (see also Chen and Eisenberg, 1993a). Briefly, the permanent charge is the charge density created by the chemical bonds of the protein, the channel contents are the ions within the channel, and the induced charge is the rearrangement in charge (in channel and pore) created by the normalized dielectric constant $\widetilde{\varepsilon}$ and the deviation $\left[\Delta(1-z)-\Phi(z)\right]$ in potential from the (spatially) constant field, namely, $\Delta(1-z)$, which has been assumed so widely in traditional physiology and channology.

The coupling between concentration and potential is most clearly seen if a single integro-differential equation is written combining the Nernst-Planck and Poisson equations. Just as the Nernst-Planck equations can be integrated once to give an expression for the flux, so can they be integrated to give an expression for the concentration of ions in the channel (equation 5.7 of Barcilon, Chen, and Eisenberg, 1992) which can then be substituted in the Poisson equation to give a single integro-differential equation for the potential, an equation uncoupled to any other. I present this expression to emphasize the complex nonlinear dependence of the potential profile on the contents of the channel, hoping thereby to discourage unanalyzed simplifications and approximations.

$$\frac{d^2\Phi(z)}{dz^2}+\overbrace{2\widetilde{\varepsilon}\left[\Delta(1-z)-\Phi(z)\right]}^{\substack{\text{Induced Charge:}\\ \text{Dielectric Effect}}}=$$

$$=\overbrace{-2\widetilde{\omega}_0(z)}^{\substack{\text{Permanent}\\ \text{Charge}}}-\lambda^2\sum_j z_j\frac{\overbrace{C_j(L)e^{z_j\Delta}\int_z^1 e^{z_j\Phi(\zeta)}d\zeta+C_j(R)\int_0^z e^{z_j\Phi(\zeta)}d\zeta}^{\text{Channel Contents}}}{e^{z_j\Phi(z)}\int_0^1 e^{z_j\Phi(\zeta)}d\zeta}$$

(20)

It would be a mathematical miracle if the potential profile $\Phi(z)$, namely the solution of this equation, were a linear function of $z$ (for example) or were independent of the trans-membrane potential $\Delta$ or bath concentrations $C_j(L)$ and $C_j(R)$. The flux and all other properties of the channel are a strong function of the potential profile, so every property of the channel is expected to be a complex function of experimental conditions.

Approximations to the electric field, or other properties of the channel, may be possible, under certain restricted conditions, but they are inherently unlikely, and therefore need to be justified analytically and numerically under the range of experimental conditions in which they are used. For example, Chen, Barcilon, and Eisenberg, 1992, analyzed five special cases (in which the permanent charge density was zero!) and show that in four of them the traditional constant field expression (Hille, 1992) can be derived. Whether the constant field expression, or any other simple expression, can arise in the presence of permanent charge is not known.

The fact that the shape of the potential profile changes with experimental interventions of many types restricts the validity of many of the traditional canons of membrane physiology. Most obviously, *a single profile of potential vs. distance cannot be used to describe a channel under a range of experimental conditions*, whatever theory (Eyring rate theory, electrodiffusion, etc.) is used at whatever level of the hierarchy to convert that profile into a prediction of current flow. Furthermore, it is not possible to guess how the potential profile changes shape as the trans-membrane potential changes. It may be that (under some conditions) it is correct to assume that the trans-membrane potential only changes the slope of a (spatially) linear component of the total potential, but one does not know ahead of time when that assumption is correct. Indeed, the *assumption is likely to break down just when phenomena become most interesting!* Many theories of the open channel are subject to this criticism. Theories of gating are also subject to this criticism if they are sensitive to the shape of the potential profile as well as the difference in potential from one side of the channel to the other.

The sensitivity of the potential profile to experimental conditions has profound effects for the analysis of all membrane transporters, not only channels. For example, it has long been believed (Hodgkin, 1951; Hodgkin and Keynes 1955a, b; reviewed in Schultz, 1980; Hille, 1992) that the unidirectional efflux of an ion (measured with radioactive tracer placed on only one side of the membrane, as described above) is independent of the *trans* concentration of the main (i.e., non-radioactive) species in continuum electrodiffusion. But if the *trans* concentration modifies the shape of the electric field, the unidirectional efflux will vary with *trans* concentration. The PNP model (Chen and Eisenberg, 1993a) shows this effect clearly and in a tantalizing way: if a channel has a uniform charge density of one sign, the unidirectional efflux decreases as *trans* concentration is increased, showing a characteristic sign of 'single filing', even though the model is of a continuum fluid. If the sign of the permanent charge is changed, the unidirectional efflux increases as *trans* concentration is increased, showing a characteristic sign of mediated transport, even though the structure involved is a channel of unchanging conformation.[44]

I believe that any consistent model of ion transport—continuum, discrete, stochastic, even a simulation of molecular dynamics—is likely to produce a unidirectional efflux that depends on *trans* concentration. The contents of the channel will vary with experimental conditions, e.g. *trans* concentration. Thus, the net charge will vary with experimental conditions (and vary significantly because the number of ions in the pore is more or less equal to the charge on the channel's wall). Because flux depends steeply (often exponentially) on the potential, the current through a channel ***in any consistent***[45] ***model*** will depend strongly on experimental conditions; in particular, the unidirectional efflux will depend on *trans* concentration.

### *Flux Ratios*

Membrane transporters have been classified by the properties of unidirectional fluxes, as just described, but even more by their ratio. The PNP model just presented allows only one ratio of unidirectional fluxes, an ideal ratio quite independent of channel properties, as we have seen. However, a wide range of flux ratios is found experimentally, and this discrepancy between experiment and theory has been viewed as fundamental, as a main justification of the usual state models of ion transport in membranes and then channels.

---

[44]It is interesting that the dependence of flux on *trans* concentration was observed in the output of a numerical integration of the PNP model long before its meaning, origin, or significance were understood. The numerical analysis of the complete model was needed to uncover this property of the system, although it could have been anticipated, had we been wise enough.

[45]precisely defined earlier in the text, on page 29.

Membrane transport has been divided into two classes, channel and mediated depending on the value of the flux ratio, for nearly fifty years.[46] Channel transport is single-file, so the flux ratio is 'less' than ideal. Mediated transport involves flux coupling and so is 'more' than ideal.

Classical active transport (e.g., the coupled movement of $Ca^{++}$ and $Na^+$ ions, Blaustein, DiPolo, and Reeves, 1991, or the coupled movement of $Na^+$ and $K^+$ ions produced by the hydrolysis of ATP in the Na/K pump, see Schultz, 1980; Hoffman and Forbush, 1983; Hille, 1989; Läuger, 1991) are examples of mediated transport. The sodium conductance (i.e., voltage activated sodium channel of nerve membrane, Hille, 1992) is an example of channel transport.

Any successful theory of transport needs to predict the flux ratios seen in experiment. And continuum theories have been (understandably) rejected in the literature (e.g., Hille, 1992) because they have predicted only ideal flux ratios (see the extensive investigations of Bass' group: e.g., Bass and Bracken, 1983; Bass, Bracken, and Hilden, 1986; Bass and McNabb, 1988).

Chen and Eisenberg (1993b) show how a simple modification of the PNP model produces non-ideal flux ratios. If a phase boundary potential $\delta_j$ (a jump in electrochemical potential) is placed in series with the ends of a channel, the flux ratio is not ideal. In that case the integrated flux expression becomes

$$
\underset{\text{Net Flux}}{\widetilde{J}_j} = \overbrace{\frac{\exp\left\{\left[-\delta_j(0)+\widetilde{\mu}_j(0)\right]/RT\right\}}{\displaystyle\int_{x_1}^{x_2}\frac{\exp\left\{z_j F\Phi_j\left(x;C_j[x]\right)/RT\right\}}{D_j(x)}\,dx}}^{\text{Unidirectional Efflux}} - \overbrace{\frac{\exp\left\{\left[\delta_j(1)+\widetilde{\mu}_j(1)\right]/RT\right\}}{\displaystyle\int_{x_1}^{x_2}\frac{\exp\left\{z_j F\Phi_j\left(x;C_j[x]\right)/RT\right\}}{D_j(x)}\,dx}}^{\text{Unidirectional Influx}} \tag{21}
$$

---

[46] ever since the papers of Ussing (1949) and Hodgkin and Keynes (1955a,b). See also Schultz, 1980; Hille, 1989; Sten-Knudsen, 1978; Jacquez, 1985; Chen and Eisenberg (1993b).

where $\delta_j(0)$ and $\delta_j(1)$ are phase boundary potentials thought to arise during the dehydration and resolvation process at the outside and inside of the channel, although the physical basis of the phase boundary potentials is not known and was not specified in Chen and Eisenberg's theory.

The flux ratio of the system with phase boundary potentials is then

$$\frac{\textit{Unidirectional Efflux}}{\textit{Unidirectional Influx}} = \exp\left\{\left[-\delta_j(0) - \delta_j(1) + \widetilde{\mu}_j(0) - \widetilde{\mu}_j(1)\right] \Big/ RT\right\}$$

$$= \quad \textbf{\textit{Ideal Flux Ratio}} \quad \times \quad \left[\exp\left\{\left[-\delta_j(0) - \delta_j(1)\right]\Big/ RT\right\}\right]$$

$$(22)$$

The flux ratio here can have any (positive real) value, depending on the difference in values of $\delta_j(0)$ and $\delta_j(1)$. That is the main result of Chen and Eisenberg (1993b), quite independent of the specific description of $\delta_j$ implemented there. Indeed, the underlying mechanism of the phase boundary potentials is not known or postulated in the theory, although some properties of the mechanism and potential are clear.

    For example, phase boundary potentials of this sort disappear when flux disappears; otherwise, current could flow in the absence of an overall driving force and thus violate thermostatics.[47] So phase boundary potentials cannot be equilibrium phenomena independent of flux through the channel. But nonequilibrium potential drops are well known to most scientists, for example, the ohmic potential drop across electrical resistors, and are found at many interfaces and phase boundaries where ions undergo large changes in state (e.g., hydration). For example, nonequilibrium potential drops are found at the interface of gases and solids (Ward, 1977; Ward and Elmoselhi, 1986), and are described by an expression closely related to the description of $\delta_j(0)$ guessed by Chen and Eisenberg. Nonequilibrium potential drops called over-potentials are found at electrodes when applied current drives them away from equilibrium. Over-potentials are

---

[47]as can be seen by setting the electrochemical potentials in both baths equal , viz., $\widetilde{\mu}_j(0) = \widetilde{\mu}_j(1)$ , in equation (21).

described by the Butler-Volmer equation (Bockris & Reddy, 1970) which is nearly the same as the analogous expression of Chen and Eisenberg or of statistical rate theory.

Chen and Eisenberg elaborated their theory to allow the flux of one species of ion to be driven by a gradient of another. If the phase boundary potentials of the two species fall in a certain broad range, this cross-coupling is compatible with irreversible thermodynamics (because in that broad range the local total entropy production, i.e., local change in temperature, is positive). The flux ratio in that system is determined by asymmetry in the phase boundary potentials. The interaction of the different ions in the dehydration/resolvation step at the ends of the channel provides the driving force for the 'active' transport.

In bulk solutions, it is well known that one ion can drag another by electrophoretic and relaxation effects, allowing the electrochemical potential gradient of one ion can drive the movement of another ion. Flux coupling is a commonplace phenomena in ordinary ionic solutions. It is "… always apparent when pronounced differences exist between the mobility of similarly charged ions." (Kortüm, 1965, p. 211). In fact, as previously mentioned, ions *in bulk solution* can in this way be induced to move against their own concentration gradient (Tyrrell and Harris, 1984, p. 377). Thus, one can easily imagine how a similar mechanism might apply at the selectivity filters at the ends of channels. There, one can imagine that the mobility of different ions is quite different (which is how selectivity might work); the electrophoretic and relaxation effects on different ions are different (with the channel protein taking the place of water: Nitzan and Ratner, 1994); and energy is transferred from the gradient of one ion to the movement of another. Despite these current speculations, I must clearly say that the physical basis of the coupling postulated by Chen and Eisenberg (1993b) is not known; an analysis with atomic resolution is dearly needed to see if the phenomena of mediated (non-ATP driven) transport might occur in a channel protein of one conformation.

*Pumping by Field Coupling*

The PNP model described up to now is what engineers call a two-terminal device, a structure in which the path for current flow has no branches. Branches are needed, however, to give many semiconductor devices (that use the same physical laws as the PNP theory) their useful functions, e.g., field-effect transistors, bipolar transistors, indeed integrated circuits in general (Horowitz and Hill, 1989; Sze, 1981; Taylor, 1987; Roulston, 1990). It is not surprising then to find that the PNP model is capable of rich behavior when it is extended to branched geometries.

In particular, the PNP model (of a channel of just one conformation) can also produce ATP driven active transport, say of $Na^+$ and $K^+$, if it is extended to the branched structures found in some channels (Miller, 1982; Matsuda, 1988; Richard and Miller, 1990; Weiss, Wacker, Weckesser, Welte, and Schulz, 1990; Weiss, Abele, Weckesser, Welte, Schiltz, and Schulz, 1991).

Imagine a **Y** shaped channel, with the top (i.e., two sided end) of the **Y** inside the cell and the bottom (i.e., one sided end) outside (Figure 1). Imagine that an ATPase is present at the end of the channel in the upper left branch[48] of the **\*Y** on the inside of the cell membrane. If that branch **\*Y** of the channel were anion selective, the phosphate ions produced by the ATPase would flow down into the channel, driven both by the concentration of phosphate (created by the ATPase by the hydrolysis of ATP) and by the local electric field. Imagine that the other upper branch of the channel (the right branch of the **Y\***) were nonselective. Then, the phosphate would come back out of the channel on the same (in)side it was created, moving through the nonselective upper right hand branch of the channel. If the bottom part of the channel (the vertical stroke of the **Y**) were cation selective, the movement of the phosphate (in the upper right branch

**Y\***, the nonselective branch of the channel) would move $Na^+$ and $K^+$ because phosphate would be the major determinant of the electric field within that part of the of the channel. In this way, the energy of ATP hydrolysis would drive the movement of cations without conformation changes in the protein. The electric field would couple the energy produced by the hydrolysis of ATP to the movement of cations. In particular, the electric field driving the movement of cations would be (mostly) determined by the phosphate concentration, in turn determined by the ATPase.

---

[48]marked with an asterisk **\***

The Na/K pump, of course, moves $Na^+$ and $K^+$ in a definite ratio (Schultz, 1980; Hille, 1989; Läuger, 1991; Hoffman and Forbush, 1983) and the model suggested so far would not produce a definite ratio of $Na^+$ efflux to $K^+$ influx. That would arise if phase boundary potentials $\delta_j$ were present as described previously in this paper (see Chen and Eisenberg, 1993b, for details). The ratio of the $\delta_j$'s at the ends of the channel (here at the bottom of the vertical stroke of the **Y** and at the top of right hand stroke of the **Y\***) would determine the flux ratio. The flux ratio could have any value (in a wide range) including the values usually found in experiments.

In this way, a structure of one conformation could pump ions, with energy derived from ATP without changing conformation. The requisite change in shape would be in the electric field. Many of the conformations postulated in traditional analysis would correspond to different shapes in the electric field, as ATP is hydrolyzed, phosphate enters the channel, etc. The occlusion phenomena so extensively studied (e.g., Forbush, 1987) would represent changes in the shape of the electric field within the channel produced by experimental interventions. Nonelectrolytes could be moved by the mechanism of dielectrophoresis in which uncharged objects move in a nonuniform electric field (Pohl, 1978).

One interesting feature of this model is the location and properties of the ATPase itself. In this model there are no special demands on the ATPase. Any one will do, any one that converts ATP to a high concentration of phosphate, and changes the local electric field in the channel. In particular, the ATPase need not have a complex structure capable of 'swallowing' one ion, disgorging it in response to another ion, then turning to swallow the other ion, etc.

Models of the Na/K pump involving a complex cycle of conformation changes have always posed a certain evolutionary (and perhaps conceptual[49]) difficulty. Because membranes are imperfect and leak $Na^+$, all cells have to pump $Na^+$ out of their cytoplasm. Otherwise, the cells will soon accumulate $Na^+$, and, even worse, water, thus increasing in volume until they burst. The incompressibility of water and the weakness of a membrane (only two lipid molecules thick) guarantee that bursting occurs quickly if the Na/K pump is blocked, in seconds in the typical sized cell (say 10 μm in diameter). For a cell to exist, it needs an active transport system, a pump, usually a $Na^+$ pump.

It is not clear how the earliest cells in evolutionary history could have evolved an enzyme as complex as that postulated in traditional models of the Na/K ATPase. Yet the earliest cells must have had ionic pumps so they could exist long enough to reproduce. The PNP model of the Na/K pump avoids this difficulty. In it almost any (nonspecific) ATPase will do, as long as it creates a high local concentration of phosphate. The special requirements are in the channel and its boundary conditions, not the ATPase itself.

Of course, this kind of evolutionary argument is more interesting than decisive. The enzymes involved in DNA replication perform complex essential tasks and arose early in the history of life. They may also undergo complex conformation changes as they function—I do not know—but it is easier to imagine[50] how this might have happened than it is to imagine how an might have evolved the complex set of conformation changes (of the traditional state model of the $Na^+ /K^+$ pump) before a membrane had evolved to define a cell.

Clearly, the three dimensional structure of the $Na^+ /K^+$ ATPase must be known before speculations of the sort I have indulged in can become real science. Nonetheless, the properties of branched channels are worth pursuing, because they exist, because they are capable of rich behavior, and because they are used for that very purpose in analogous systems in semiconductors. Indeed, branched geometries form the solid state technology of contemporary electronics.

---

[49]See footnote 39.

[50]at least for me, although admittedly this lack of imagination may reflect my limitations more than those of evolution.

## GATING IN CHANNELS OF ONE CONFORMATION

So far, I have only discussed properties of the open channel. Little has been said of how the channel opens and closes. And little can be said in general, if the channel opens and closes by a conformation change, that is to say, by a substantial movement of many atoms of the channel protein. Conformation changes cannot be understood without structural data—including measurements of structure as a function of time—and so they cannot be analyzed by *a priori* physical theory. They cannot be reduced to atomic biology, unless the laws of motion of proteins are known. Not only are such laws not known, it is not clear that they exist in any form simpler than the molecular dynamics of all the atoms of the protein. Different conformation changes may require different descriptions. Macroscopic rules may not be able to describe even one conformation change, under a range of conditions let alone all the conformation changes of all proteins.

Open channels of one conformation have certain properties that hint of gating, however, and it is important to mention and discuss these, albeit in a speculative manner.[51] Current flow through an open channel can be controlled in a way that does not involve a conformation change at all, that does not involve a change in shape of the channel protein, but does involve a change in the shape of the electric field within the protein's pore (Chen and Eisenberg, 1992).

If the shape of the electric field in a channel protein changes, the current flow through the open channel changes. If the shape of the electric field changes a little, the current flow can change a lot. For example, if a barrier height increases by a few $kT$, the current flow would decrease substantially and the channel would appear to close, at least in the sense that no detectable current would flow through it. As we have already discussed, the shape of the electric field depends on many variables, for example, on the concentration of ions in the bath, on the trans-membrane potential, and on the potential across the membrane far from the channel's pore, or even away from the channel protein itself. Thus, changes in any of those variables could close or open the pore of a channel, just by changing the shape of the electric field, without changing the conformation of the channel protein, just as the potential on the gate of a field-effect transistor changes the shape of the electric field in its conducting 'channel' and thus changes its conduction. In a similar way, gating of a channel could be produced by a change in membrane potential, or by the binding of an agonist to a nearby receptor (which changes the far-field), all without a change in the conformation of the protein, as that word is ordinarily understood. Of course, something is changing shape when the electric field changes shape, namely the distribution of charge and, if one wished, one could call this a conformation change. But that is not at all what enzyme kineticists have in mind when they speak of conformation change: they mean a change in shape of the protein. They mean a geometrical or steric effect, not an electrical effect. And changes in geometrical shape and changes in electric field are not equivalent, particularly when constructing theories. A conformation change of an electric field is much more tightly constrained by known physical laws than a conformation change in the steric sense. Changes in the electric field are described by specific known physical laws (the laws that relate potential to charge, in continuum models this is Poisson's equation; in discrete models, this is Coulomb's law summed over all charges) whereas a shape change in a protein cannot be described in anything like such a specific way. Calling a change in shape of an electric field a 'conformation change' might make the idea more palatable by connecting it with an already accepted notion, but it would obscure the essential difference between two quite different mechanisms. I reserve 'conformation change' for its original use.

---

[51]entirely speculative because a quantitative theory of the gating of a specific channel has not yet been constructed, let alone tested, although we are trying, so far with little success.

The kind of gating just described, resulting from a change in shape of the electric field, but not a change in shape of the distribution of permanent charge, is well known in semiconductor physics. In semiconductor devices, like transistors or thyristors, changes in current are produced by a change in the electric field (Shur, 1990; Seeger, 1991; Mock, 1983; Markowich, 1986; Hess, 1988; Jerome, 1994). They are never produced by a change in structure, that is to say by a change in what we call the distribution of permanent charge or what the semiconductor physicists call the doping function. Changes in the electric field produce the enormous richness in behavior seen in semiconductor circuits. Indeed, a computer can be built entirely from (three terminal) devices made of semiconductors obeying (two dimensional) drift-diffusion equations,[52] and so changes in the electric field can produce all the behavior a computer can produce, ***including all the behavior that can be coded in a semiconductor memory***. In this way, a two dimensional PNP equation with complex distribution of permanent charge can reproduce all the information and behavior human beings can code into words or numbers! No wonder a general theory of the properties of the PNP equation (with arbitrary charge distribution) looks hard to derive (Jerome, 1994).

Channels are thought to be essentially one-dimensional devices, with the exception of the branched channels are known as we have discussed. Putting those aside, for the moment— although they might have interesting gating behavior as we discuss later—the question arises whether simple unbranched one-dimensional channels of one conformation, described by the PNP model, are capable of the richness of behavior seen in semiconductor devices. In particular, can such simple structures gate the way thyristors do (Sze, 1981; Taylor, 1987)? Can channels open and close (conduct or block current) as semiconductor devices do, simply by changing their electric field without changing their arrangement of permanent charge? I would dearly love to answer these questions, but cannot. We know that the PNP equations have this property (Chen and Eisenberg, 1992) as discussed later in this paper, and we know that channel gating (or some types of channel gating) *might* work this way, but we do not know if it actually does. It seems likely to me that some of the phenomena of gating arise this way, either subconductance states, flickers, or the spontaneous or controlled opening of channels. But evidence does not force this conclusion, only intuition (e.g., prejudice) motivates it. Until we can construct a discrete PNP theory, in which charges are discrete but the potential profiles are computed (not assumed), we cannot tell whether charge rearrangement might produce some or all of the phenomena of gating seen in single channels.

Some general conclusions can be reached without a discrete theory. The sensitivity of the electric field in the channel to the contents of the channel's pore has some interesting consequences that will survive the details and analysis of any consistent theory. For example, it is clear that the contents of an open channel have something like the same total charge as the wall of the channel: otherwise, the un-neutralized charge would create large fields. These would either be large enough to destroy the three-dimensional structure of the channel protein (which is after all held together by much the same electrical forces that act between permeating ion and channel wall), or, in a less extreme case, they would create insurmountable barriers to current flow, potential barriers that would produce the closed channel seen experimentally, even though the pore remained physically open.

---

[52]nearly the same as the PNP equations in a circuit of branched channels, differing only in the induced charge term.

Indeed, the sensitivity of the electric field to channel contents is so great that one can reach a rather important general conclusion about single channel currents just from their existence, as rectangular events with one height (but variable duration). ***Single channel currents of (much) the same amplitude probably arise from channels with (nearly) the same potential profile and thus with (much) the same contents***.

If the contents of the channel changed in any material way, the current flow would change much more. The change in contents would produce a change in the net charge of the channel protein (i.e., a change in the sum of the channel contents plus the charge on the channel wall) which would produce a change in the electric field. That change in field would produce a larger change in measured current because flux is a sensitive, usually exponential function of barrier height. The fact that single channel currents do not change from opening to opening (more than the noise in their measurement), nor do they drift during the duration of their opening, implies that the (time averaged) shape of the electric field and the (time averaged) contents of the channels do not drift during an opening, or from opening to opening. If the channel contents did change significantly, the electric field would change shape, most likely producing a large barrier somewhere. A barrier anywhere is enough to prevent conduction (since the channel is a series device). Thus, ***any significant change in contents of a channel is likely to close it!***[53]

---

[53]Some changes in contents would not change the shape of the electric field. Examples would include (1) the exchange of one $Na^+$ for another; (2) the exchange of one $Na^+$ for another isotope of $Na^+$; or (3) the exchange of one $Na^+$ for another univalent cation with indistinguishable interactions with the channel protein, if such exists.

### *Gating by Field Switching*

An intriguing clue to a possible mechanism of gating is given by the mechanism of switching in a thyristor (Sze, 1981; Taylor, 1987). The thyristor is a semiconductor device with (typically) two positive and two negative regions of doping (what we call permanent charge), more or less two transistors back to back, not quite a two terminal device. The thyristor functions by switching between conducting and non-conducting states. The thyristor has two different current-voltage relations, one conducting and the other non-conducting, namely isolating or insulating. Both current-voltage relations are stable, that is to say, once the device has switched from one state to the other, it can stay put (in conducting or non-conducting state) over a wide range of conditions.

It is intriguing to examine unbranched channels with charge distributions similar to those of a thyristor (Chen and Eisenberg, personal communication). It is a particularly seductive line of thought because Numa has suggested (Noda *et al,* 1984) that the classical voltage activated sodium channel has a shockingly similar distribution with four maxima in its distribution of permanent charge, although evidently he was unaware of the analogy with thyristors. If the PNP model is used to compute the current-voltage relation of Numa's distribution of charge, the current-voltage relation is not unique (Chen and Eisenberg, 1992). There are (at least) two current-voltage relations, one with a tiny average slope (i.e., low conductance), and the other with a substantial slope, of something like 10 pS. These different states of the channel look quite like closed and open sodium channels.

The 'closed' channel has a distribution of potential (computed from the Numa arrangement of permanent charge) that has much the same shape as the distribution of permanent charge, with four maxima. In the closed state, almost no mobile (ionic) charge is present within the channel; thus, the permanent charge is not neutralized and the potential barriers are large. Current cannot flow because the channel contains little mobile charge and that charge cannot cross the barriers. As one might expect, the current-voltage relation of the closed state has very small average slope (some femto-siemens or less), but the relation is quite nonlinear. If such a state exists, it might contribute to the so-called background (or seal or leakage) current found in patch clamp experiments. One might expect part of that ***background*** current to be removed by pharmacological agents, i.e., by channel blockers. Effects of this sort have in fact been observed in the closed porin channel and in background currents in preparations of potassium channels (Fred Quandt, personal communication).

The 'open' channel computed from the Numa charge distribution has very different properties. Its distribution of potential is monotonic, showing no signs of the maxima of the permanent charge distribution, or of the closed channel, because the open channel contains substantial amounts of neutralizing charge, namely something like one or two ions (smeared into a continuum in the PNP model, regrettably but inevitably). This neutralizing charge screens the permanent charge of the channel, producing a potential profile without barriers. The resulting current-voltage curve is quite linear (because the potential profile is monotonic) and with relatively large slope, i.e., conductance. The channel is quite filled with charge and that charge has no barriers to cross. In this way, a single unbranched structure of permanent charge could account for both the open and closed channel properties of a single channel, without any change in conformation in the normal sense of the word.

Of course, the electric field in such a model changes shape dramatically as the channel opens and it is that change which allows current through channel. The charge that changes the shape of the electric field includes the charge in the channel, the charge induced in the channel protein, and the (nonlinear) induced charge at the boundaries of the system. The very first charge that enters the channel after a step change in voltage 'fights its way' against the potential barriers of the closed channel. This takes a long time[54], as one might expect, but eventually enough charge enters the channel to screen and shield the permanent charge, lowering the potential barriers, allowing more current to flow, creating the open channel. The ON response (from an empty, unshielded, low conductance state, to a filled, shielded, high conductance state) is thus S-shaped, starting slowly; interestingly the OFF response (from a filled, shielded, high conductance open state to an empty, unshielded, low conductance closed state) is monotonic, starting very quickly. This time course is quite like current through an ensemble K channels of squid nerve, although the time scale we compute is too fast.

---

[54]microseconds in the unpublished simulations of Chen and Eisenberg (1992) of the time dependent PNP model containing just univalent ions.

What is not clear is whether this kind of field switching actually occurs in biological channels, as well as in the PNP theory. *If* field-switching does occur, however, it could produce many of the phenomena of gating (e.g., the voltage or agonist driven time dependent gating analyzed by Hodgkin and Huxley, 1952, or Katz, 1969), the opening and closing of single channels, or their flickering and switching into subconductance states, as Duan Pin Chen and I realized as soon as we constructed the PNP model (Chen and Eisenberg, 1992). But, to be fair, I must plainly say that we did not know then, and do not know now, if such field-switching occurs at all in real biological channels, as appealing as the idea is, at least to me.

### *Gating Current*

The changes in contents of channels just described are not electrically silent. If the charges inside a channel move, an exactly equal movement of charge occurs in the surrounding solution (according to Maxwell's equations) and the external charge movement can be collected in electrodes and recorded by electronics, if they are faithful enough. If a rectangular pulse of voltage is applied to such a channel, the charge flowing on the ON of the pulse will equal the charge flowing on the OFF, because the channel contents switch from one state to another and then back to the original state. A plot of this charge vs. voltage will be nonlinear, because the dependence of channel contents on potential is nonlinear.

Nonlinear displacement current with these characteristics was postulated by Hodgkin and Huxley (1952) and found by Schneider and Chandler (1973), Armstrong and Bezanilla (1974) and Keynes and Rojas (1974). They attributed the current to the movement of a voltage sensor, presumably a part of the channel protein (more recent references in Hille, 1992). This kind of gating current might be called allosteric current or conformation current, because it results from the change in shape of the protein.

Changes in the contents of a channel might mimic some or all of the properties of an allosteric current, of nonlinear gating charge. The voltage sensor might be just the contents of the channel; in any case the contents of the channel are likely to be part of the voltage sensor.[55] Clearly only a few ions can fit into a channel (at one time), and so one would expect a gating charge of only a few elementary charges in this model of gating current, as is found experimentally. This finding of small nonlinear gating charge in a channel protein is hard to understand if the current arises from a large scale conformation of the protein, as is commonly believed. The channel protein contains thousands of polar chemical bonds with partial charges of 0.1-0.6 elementary charges (Schulz and Shirmer, 1979), and so it is hard to see how a conformation change involving so many (electrically charged polar) chemical bonds can produce as *little* (nonlinear) gating charge as is seen experimentally.

On the other hand, a field-switching theory of gating current has serious difficulties dealing with some of the other properties observed experimentally. The slow time course of gating current might not be expected from a change in the contents of a channel. The time dependent Nernst-Planck equation changes contents in nsec, if it is analyzed assuming a constant electric field (Cohen and Cooley, 1965; Cole, 1965, 1968), independent of transmembrane potential and time. If the channel supplies the charge to modify the field, as in the time dependent PNP model, the channel contents change much more slowly, but still in **micro**-seconds at the longest[56] compared to the **milli**-seconds seen experimentally.

The field-switching theory has at least one more difficulty. It implies that the shape of gating current should depend on the species of ion which enters and then exits the channel, thereby creating the (equal) ON and OFF charge movement. Different ions should penetrate the channel differently even if they do not permeate. Experiments, however, show much the same gating current no matter what impermeant ions are present.

---

[55]How could they not be? The charge in the channel's pore is certain to move in response to a change in the electric field and it is also likely to control (at least in part) the flow of current in the pore, at least in unbranched geometries (see later discussion). The issues are only whether this charge movement is actually used for biological purposes and whether it occurs slowly enough to be measured by the usual protocols and equipment.

[56]Only univalent ions were present in these unpublished simulations (Chen and Eisenberg, 1992). It is possible that $Ca^{++}$ ions would have as large an effect on the theory as they do on real channels and so the theory and calculations need to be reformulated to include them.

In view of these significant problems, I believe that change in the charge within an unbranched channel cannot account for the gating current seen experimentally, probably because that kind of charge moves too quickly to resolve experimentally.

### Gating in Branched Channels

The discussion of gating in this paper, and all the literature I know, has assumed that the classical voltage sensitive channels have an unbranched structure. It is worthwhile speculating briefly how a branched, **Y** shaped channel might behave (Miller, 1982; Matsuda, 1988; Richard and Miller, 1990; Weiss, Wacker, Weckesser, Welte, and Schulz, 1990; Weiss, Abele, Weckesser, Welte, Schiltz, and Schulz, 1991). If one branch of the channel were unselective, say the left branch of the **\*Y**, but separated by a high potential barrier from the rest of the channel, it and its contents would help control the potential in the main branch of the channel. The contents of the left branch in that case would be the gating charge that moves in response to changes in the trans-membrane potential, thereby controlling the field and permeation in the main branch of the channel. If the **Y** were sufficiently broad, the voltage sensor branch **\*Y** would sense the membrane potential, quite independently of the potential within the main branch **Y** of permeation. Thus, permeation would be controlled by the membrane potential (chiefly) and not be sensitive to the current flow and potential within the permeating path itself. The fact that permeation through voltage sensitive channels is a function of the trans-membrane potential (chiefly) and is quite insensitive to current flow through the channel has been known for some time (Cole, 1947; Hodgkin and Huxley, 1952) and was one of the motivations for the original design of the voltage clamp. Any satisfactory model of gating, with branched or unbranched channel, will predict gating phenomena quite insensitive to the amount of current flowing through the channel. But anything that changes the current flow through the channel is likely to change the shape of the electric field within the channel's pore. And anything that changes the shape of the electric field in the pore is likely to change gating that occurs there (e.g. gating by field switching in the pore). Thus, a successful theory of an unbranched channel will somehow isolate the properties of gating from the current flowing in the channel's pore, probably by making gating quite insensitive to the shape of the electric field within the pore.

A branched channel, as just described, would have properties quite similar to a field-effect transistor, in which (gate) potential (chiefly) controls current flow. The branched channel described previously as a Na/K pump would have properties quite similar to a bipolar transistor, in which both (base) current and potential control current flow. It is appealing, but wildly speculative at this stage of knowledge, to imagine that evolution might have constructed two devices so like the two main types of transistors, using them to performs its functions as we use them to make our machines.

### *Blocking*

The effects of ions within a channel are hardly a surprise to workers who have studied the blocking effect of slow ions (i.e., weakly permeable ions) on open channel currents (Hille, 1992, summarizes this large literature). They know that a channel can be closed (i.e., blocked) by what appears to be the entry of a single ion into the channel's pore. Perhaps the entering ion simply changes the electric field within the channel, without necessarily changing the shape of the protein (although of course that might happen as well). The ion need make a rather small change in the shape of the electric field to block current flow because the flow tends to be an exponential function of peak barrier height. A change of barrier height of 4.6 $kT$ would change flux by a factor of 100 and produce a 'blocked' channel.

In a way, what I suggest is simply a renaming of a 'blocked channel' to a 'channel with an altered electric field'. The unspecified blocked state is now identified with a state of a different distribution of charge and thus potential. But more is involved than renaming. The distribution of charge and potential are governed by specific physical laws[57] and so the new description can be specific, predictive, and falsifiable. It is also consistent. None of those features characterize state models of the usual type.

The entry of an additional ion, adding charge to the usual contents of the channel, clearly should change the potential profile substantially in any consistent model, likely producing a blocking event. But what if one type of ion exchanges with an ion of another type, say a monovalent cesium replaces a monovalent potassium? The charge does not change, so why should the field change? Clearly if the ions are indistinguishable, the field will not change. If the channel is nonselective between cesium and potassium, it cannot tell them apart and the field will not change, nor will permeation, nor will single channel current. The open channel current will be maintained and there will be no obvious experimental signature of the new contents of the channel.

---

[57]Poisson's equation and the Nernst-Planck equations in the PNP model; Langevin equation in the stochastic models; Coulomb's law plus the molecular dynamics of protein, water, and permeating and blocking ions in the highest resolution models

But cesium and potassium are not the same. They have quite different atomic radii and so quite different electric field strengths at their edge where they interact with water and (say) carbonyls of the channel protein. Their interactions with other atoms are described by quite different 'potentials' in the simulations of molecular dynamics and overall energies of interaction will differ by several $kT$.

Thus, the various properties of the permion would be expected to change. The diffusion constant, the potential barrier, the effective mass would all change. A PNP theory of this kind of blocking would require additional information describing the modified conformation of the channel, information that might come from calculations of molecular dynamics. Analysis of this kind of blocking would require an explicit link between the hierarchies of models because the conformation of the channel protein is expected to be different when it is occupied by a cesium rather than a potassium. The movement of the atoms of the channel protein will change the distribution of permanent charge, even though no covalent bonds have been changed. The existing permanent charge will be in different places, in different densities (if one chooses the continuum point-of-view). Induced charge will be different as well. In this case, one would expect the channel protein to be in a different state, as it is usually said to be. That state can be specified in the physical language of atomic biology by theoretical or (better) experimental analysis. Structural determination by x-ray crystallography or NMR can (in principle) give all the atomic coordinates, but those techniques cannot yet be routinely applied to channels. Numerical simulations of molecular dynamics can give a theoretical prediction of the new structure (with cesium in the channel's pore) and that is what we will work with for the foreseeable future, until structures are experimentally known.

## BACK TO THE HIERARCHY: LINKING LEVELS

The consequences of contents/field coupling in the macroscopic PNP model are so profound that we have nearly lost sight of where we started earlier in this paper, in an hierarchy of models of which the PNP is the most lofty, but with lowest resolution and therefore most in the air, most remote from the solid foundation of molecular dynamics! The properties of that model are important and exciting, particularly as they reveal the physics needed in any consistent model at any level of resolution, but the PNP model needs to be related to models with higher resolution, those resting more solidly on the foundation of a hierarchy, models with atomic resolution, like molecular dynamics.

The proper way to proceed when linking levels of the hierarchy is to actually derive, by mathematics alone, the dynamics, conservation laws, and boundary conditions, of the low resolution model from the high(er) resolution model. That derivation will show explicitly where the laws come from, what the functions and parameters mean, and how the boundary conditions relate to the experimental situation. Such derivations are rarely possible. In the problems I have worked on, we could only do this in one case (Eisenberg, Klosek, and Schuss, 1994): there, it took nearly eight years to derive the flux laws of the Nernst-Planck equation from the stochastic properties of the underlying diffusion trajectories. And there it was necessary to assume the electric field and consider only conditions in which the electric field does not change. So, it is not clear how long it will take to derive a coupled macroscopic model—a PNP equation—from a description of stochastic motion of individual ions, let alone from molecular dynamics in atomic detail. Until then, it is difficult to discuss even the meaning (in terms of parameters of discrete high resolution models) of the effective parameters of the smoothed low resolution models. But at least one important thing can be said about one of the macroscopic variables of the PNP model. The effective concentration of an ionic species in a channel depends as much on the duration of time a channel is occupied by the ion as it does on the number of ions that can fit in (a given volume of) the channel.

***Occupancy.*** A channel is not always occupied by the same ions. It may even spend a significant fraction of time occupied by no ions. The time that the channel is occupied by each ion (or a particular combination of ions) is a variable, dependent on all the parameters of the problem and so the effective concentration averaged over all times depends strongly on the duration of each occupancy state of the channel. Thus, a theory that links stochastic theory with the PNP model must determine the duration of each occupancy state, and how that duration varies with parameters and experimental conditions. Indeed, in a time dependent problem, the theory must predict the time dependence of the occupancy states as well, e.g., how they vary after a step in potential.

Some of this theory can be avoided if the single channel current represents the only conducting occupancy state[58], because, as argued previously, any other contents of the channel are likely to produce a large internal potential barrier that blocks current flow. If we accept that argument—or better yet, if molecular dynamics or experimentation shows it to be correct—the various statistics of the open state (e.g., the open probability, histograms of open times, and so on) *directly estimate* the properties of a single occupancy state, making the theory quite specific by dramatically reducing the number of unknown parameters. Furthermore, in this situation occupancy and gating become intimately linked, even identical (in some reduced cases). Changes in occupancy produce individual channel opening and closing; changes in occupancy are produced by random motions of ions over the ***pre-existing*** potential barrier (that present just before the ion enters the channel and changes the occupancy state). The pre-existing potential barrier would be a function of trans-membrane potential, ionic conditions and so on, and in that way the qualitative properties of gating might arise ***as a result*** of changes in occupancy. Unfortunately, however, the measured statistics of the open channel are not in themselves enough to imply a theory of occupancy let alone gating. They do not say what is in the channel during its open time or how much of it is there, so a full analysis is not possible, without further experimentation to determine the number and type of ions in the channel, and, of course, a consistent theory of occupancy (i.e., a theory of ion entry, ionic movement within the channel, and ion exit, on one side or the other). The theory must predict (not assume) the dependence of occupancy on the various parameters of the system if it is to be useful.

---

[58]See footnote 53.

Changes in occupancy may produce many of the more complex phenomena of channels: flickering and subconductance states both are easy to explain as the consequence of a change in occupancy. Flickering would result from a channel occupied by an ion that raises potential barriers enough to cut off current flow. Subconductance states would come from channels occupied by an ion that hardly changes the potential profile, that is to say, that changes it just enough (say a few kT) to make a small (say 10%) change in conductance.

Occupancy affects each of the parameters of macroscopic smoothed theories like the PNP model. Each needs microscopic interpretation. Even parameters like the density of permanent charge are likely to depend on channel occupancy (Chen and Eisenberg, 1993a: Discussion Section) and general analysis is difficult because the various effective parameters (potential, friction, permanent charge, mobile charge, etc.) are not averaged the same way. They do not depend on occupancy in the same way. It seems unlikely that verbal analysis will suffice. A real theory is needed.

The problems just described occur when we try to link macroscopic with microscopic, smoothed with stochastic. But similar problems arise whenever one level of the hierarchy is connected to another.

Every theory has its effective parameters. The two body 'potential functions' of molecular dynamics are the output of quantum chemical considerations and calculations. The 'potential functions' of the Langevin theory are the output of molecular dynamics and perhaps macroscopic theories as well (because of the long range nature of the electric field discussed previously). The input of one theory is the output of the other. Linking input and output is an important part of building an hierarchy extending from atomic structure to biological function. I only wish we knew how to do it.

So we proceed as best we can, inventing methods as we proceed, stumbling on generalizations by working on the particular. Peeking through the keyhole of an open channel, we gaze at the horizon of atomic biology in general.

## IS THERE A THEORY?

With all the complexity of the hierarchy and the difficulty of linking its levels, it is easy to become discouraged. After all, how can I be so sure there is a theory, simpler than the molecular dynamics of all the atoms of the protein, that describes the function of a channel?

The reason I am so sure is evolution. It is hard to believe that life could exist if its main molecular functions did not follow reproducible, reasonably general laws. An engineer requires his amplifier or logic element to follow simple deterministic rules under a range of conditions. Evolution requires its molecules to perform the same way. They function reproducibly under varying conditions to perform life's functions. The engineer's amplifiers are ***internally*** governed by a hierarchy of models at least as daunting as the hierarchy of channel models, but ***externally*** they follow the simple phenomenological rules of circuit elements. Otherwise, the engineer would not know what they would do next and he could not design a functioning system.

What are the biological equivalents of these linear amplifiers and nonlinear logic elements? What are the building blocks of channel function? What are the phenomenological rules that determine the biological function of a channel, or protein, for that matter? We do not know; indeed, finding these rules is one of the main goals of modern biology. We can be confident a simple rule exists (for the natural function of a biological molecule), but ***we do not know at what level of the hierarchy it will emerge.*** The biological function might be the result of atomic interactions (e.g., selectivity) and so need to be described at the subatomic level of molecular orbitals. The function might be the result of twists in the permeation path, and so need to be described at the level of molecular dynamics. The function might be the result of interactions of individual ions and so need to be described at the level of stochastic dynamics. Or it might be the result of the macroscopic electric field and be at the smoothed level of continuum theories. We do not know at what level of the hierarchy to work until we find the simple phenomenological law that describes a natural function. Then we know. But then it is too late to avoid a great deal of (seemingly unnecessary) work on 'wrong' levels of the hierarchy.[59]

---

[59]Perhaps practical discoveries will arise serendipitously from the 'unnecessary' work, as they usually have in the past.

***At what level will the adaptation be found?*** It is unfortunately necessary to examine every level of the hierarchy of models linking structure and function until we find where and how evolution created an adaptation to solve its problem, to enhance reproduction, to ensure fitness, or survival. If we examine only one level of the hierarchy, we will not be able to observe or resolve some adaptations and phenomena at all, i.e., adaptations that are best understood on other levels of the hierarchy, not clearly seen from the one level we are examining. For example, at the level of continuum models we cannot describe adaptations and phenomena that depend on discrete atoms; on the level of molecular dynamics we cannot compute, and can hardly describe or observe the macroscopic electric field. Only when we look at the appropriate level of the hierarchy—the one at which the adaptation occurs—can we expect simplicity. If we observe on only one level, and that happens to be a wrong one—a level that does not reveal and resolve the adaptation—then we will be confused. We will most likely find too many phenomena that we cannot quite explain or we can explain, but not uniquely. It will be easy to become confused, discouraged, or even worse to be misled that knowledge is greater than it really is.

For these reasons, work is needed on all levels of the hierarchy until we learn where each biological phenomena fits. On that level, each biological function will be best (and most simply) described. Different adaptations will be best described on different levels, but each will follow a simple law when described on its own level, I believe.

***Simplicity, Evolution, and Natural Function.*** Simple laws can be found on some level of the hierarchy, of course, only if they exist. The argument that simple rules can be found is essentially evolutionary and depends on life's need for simple, reproducible reasonably general function in the face of a varying environment. That is a homeostatic argument on a molecular scale.

The argument does not apply, however, to all properties of channels or proteins, only to those that are directly involved in the natural function of the molecule, the function that evolution adapted and selected. Properties that are not natural functions may be complex special cases, quite changeable as the environment changes. An engineer can construct a limiter (i.e., threshold detector) from a linear amplifier by putting it in an 'unnatural' state, by changing its connections and supply voltages. The jury-rigged limiter will work, but it is likely to have hysterisis and sensitivity to noise that make it hard to use and study. Devices designed as limiters in the first place will be easier to use and study.

For the same reason, the ***natural functions of a protein will probably be (much) easier to study than unnatural ones***. Studying un-natural function is likely to be an unrewarding task. Studying voltage dependent gating in a channel that never sees a changing voltage in life (if such exists) is likely to reveal an experimental mess, a mass of hard-to-reproduce phenomena sensitive to the details of the experimental conditions and source of the preparation. Studying that channel is unlikely to reveal the general laws of voltage gated channels.[60] Phenomenological laws— general rules of natural function—will exist (I believe) describing natural functions under the range of natural conditions, but they will not apply to every experimental condition we can create.

Seeking and understanding the general rules of natural function is the task of atomic and molecular biology. Hopefully, the reader will not be too discouraged to continue what we have barely begun. Our successors will be able, I trust, to compute the natural function of channel proteins from their structure, thus creating an atomic theory of ionic channels, a small, perhaps initial piece of the eventual atomic biology of proteins.

---

[60]unless the channel had evolved from (or into) a voltage dependent channel and has a vestigial (or nascent) voltage sensor and/or voltage controlled gate!

## AFTERWORD

Biology has always been a descriptive science, as it must be, because it studies the particular outcome of a chaotic process, and is essentially an historical science studying an evolving process at just one time. Molecular biology should be seen in that light, as (for the most part) the identification, description, manipulation and control of the proteins life uses. Most biologists are needed for those tasks. Their work will change our lives more than we can easily imagine, by improving our technology in general and our health in particular. Their work will soon give us staggering control over the human life cycle.

This paper describes another kind of biology, the analytical kind, and arises from the reductionist tradition in which I was trained, by John Edsall and Andrew Huxley, Alan Hodgkin (mostly by example) and Julian Cole, among many others. I hope I have not misunderstood their lessons.

## ACKNOWLEDGMENT

An hierarchy of models is many more than any one person, certainly I, can build or master. The builder responsible for each level of the hierarchy requires a career of expertise in the techniques, traditions, successes, and failures of the corresponding physical science. I have had the fun and honor of working with many such builders, sharing the excitement of the hierarchy with them: Victor Barcilon, Ron Elber, Joe Jerome, Mark Ratner, Zeev Schuss, with the continuing help of Duan Pin Chen, who has worked with me on so many levels.

Victor Barcilon, Joe Blum, Tom DeCoursey, Truman Esmond, Lane Niles, and Mark Ratner were kind enough to make vigorous criticisms of the manuscript. I thank them for their interest, time, and help!

The steadfast encouragement and support of Andrew Thomson and the National Science Foundation made this work possible. I am most grateful.

## *REFERENCES*

Allen, M.P. and Tildesley, D.J. (1989). ***Computer Simulation of Liquids***. Clarendon Press, Oxford.

Anderson, H.L. and Wood, R.H. (1973). Thermodynamics of aqueous mixed electrolytes. *In: Water*. F. Franks (ed), Plenum Press, New York.

Armstrong, C.M., and Bezanilla, F. (1974). Charge movement associated with the opening and closing of the activation gates of the Na channels. J. Gen. Physiol. 63: 533-552.

Arnold, L. (1974). ***Stochastic Differential Equations: Theory and Applications***. John Wiley & Sons, New York.

Bader, R.F.W. (1990). ***Atoms in Molecules. A Quantum Theory.*** Oxford, New York.

Bai-Lin, H. (1984). ***Chaos***. World Scientific, Singapore.

Barcilon, V. (1992). Ion flow through narrow membrane channels: Part I. SIAM J. Applied Math 52: 1391-1404.

Barcilon, V., Chen, D.P. and Eisenberg, R.S. (1992). Ion flow through narrow membranes channels: Part II. SIAM J. Applied Math 52: 1405-1425.

Barcilon, V., Chen, D., Eisenberg, R. and Ratner, M. (1993). Barrier crossing with concentration boundary conditions in biological channels and chemical reactions. J. Chem. Phys. 98: 1193-1212.

Bard, A.J. and Faulkner, L.R. (1980). ***Electrochemical Methods***. John Wiley & Sons, New York.

Bass, L. and Bracken, A.J. (1983). The flux-ratio equation under nonstationary boundary conditions. Mathematical Biosciences 66: 87-92.

Bass, L., Bracken, A. and Hilden, J. (1986). Flux ratio theorems for nonstationary membrane transport with temporary capture of tracer. J. Theor. Biol. 118:327-338.

Bass, L. and McNabb, A. (1988). Flux ratio theorems for nonlinear membrane transport under nonstationary conditions. J. Theor. Biol. 133: 185-191.

Berg, H.C. (1983). ***Random Walks in Biology.*** Princeton University Press, New Jersey.

Berry, R.S., Rice, S.A., and Ross, J. (1980). ***Physical Chemistry.*** John Wiley & Sons, New York.

Billingsley, P. (1986). ***Probability and Measure*** , (2nd Edition), John Wiley & Sons, New York.

Blaustein, M.P., DiPolo, R. and Reeves, J.P. (1991). ***Sodium-Calcium Exchange***. Annals of the New York Academy of Sciences, Vol. 639, New York.

Bockris, J. and Reddy, A. (1970). ***Modern Electrochemistry***. Plenum Press, New York, pp. 1-1432.

Böttcher, C.J.F. (1973). ***Theory of Electric Polarization, (2nd Edition),*** Elsevier, New York.

Boyer, C.B. (1949). ***The History of Calculus and its Conceptual Development*** (Dover Publications, N.Y.)

Bray, W.C. and Hunt, F.L. (1911). The conductance of aqueous solutions of sodium chloride, hydrochloric acid, and their mixtures. J. Am. Chem. Soc. 33: 781.

Brooks, C.L., Karplus, M. and Pettitt, B.M. (1988). ***Proteins: A Theoretical Perspective of Dynamics, Structure and Thermodynamics***. John Wiley & Sons, New York.

Brumer, P. (1988). Chaos and reaction dynamics. *In:* ***Evolution of Size Effects in Chemical Dynamics. Part 1***. I. Prigogine, S.A. Rice (eds). John Wiley & Sons, New York, pp. 365-439.

Brush, S.G. (1986). ***The Kind of Motion We Call Heat***. North Holland, New York.

Burkert, U. and Allinger, N.L. (1983). ***Molecular Mechanics.*** American Chemical Society. Washington.D.C.

Cajori, F. (1980). ***A History of Mathematics,*** 3[rd] Edn., Chelsea Publishing, N.Y.

Chapman, S. and Cowling, T.G. (1970). ***The Mathematical Theory of Non-Uniform Gases***. (3rd Edition). Cambridge University Press, Cambridge.

Chen, D.P., Barcilon, V. and Eisenberg, R.S. (1992). Constant field and constant gradients in open ionic channels. Biophys J. 61:1372-1393.

Chen, D.P. and Eisenberg, R.S. (1992). Exchange diffusion, single filing, and gating in macroscopic channels of one conformation. J. Gen. Physiol. 100:9a.

Chen, D.P. and Eisenberg, R.S. (1993*a*). Charges, currents and potentials in ionic channels of one conformation. Biophys. J. 64: 1405-1421.

Chen, D.P. and Eisenberg, R.S. (1993*b*). Flux, coupling and selectivity in ionic channels of one conformation. Biophys. J. 65: 727-746.

Chiu, S.W. and Jakobsson, E. (1989). Stochastic theory of singly occupied ion channels. II. Effects of access resistance and potential gradients extending into the bath. Biophys. J. 55: 147-157.

Cohen, H. and Cooley, J.W. (1965). The numerical solution of the time-dependent Nernst-Planck equations. Biophys. J. 5: 145-162.

Cole, K.S. (1947). *Four Lectures on Biophysics*. Institute of Biophysics, University of Brazil, Rio de Janiero.

Cole, K.S. (1965). Electrodiffusion models for the membrane of squid giant axon. Physiol. Rev. 45: 340-379.

Cole, K.S. (1968). *Membranes, Ions, and Impulses*. University of California Press, Berkeley, 569 pp.

Conway, B.E., Bockris, J.O. and Yeager, E. (1983). *Comprehensive Treatise of Electrochemistry*. Plenum Press, New York.

Cooper, K.E., Gates, P.Y. and Eisenberg, R.S. (1988*a*). Surmounting barriers in ionic channels. Quarterly Review of Biophysics 21: 331-364.

Cooper, K.E., Gates, P.Y. and Eisenberg, R.S. (1988*b*). Diffusion theory and discrete rate constants in ion permeation. J. Membr. Biol. 109: 95-105.

Cooper, K., Jakobsson, E. and Wolynes, P. (1985). The theory of ion transport through membrane channels. Prog. Biophys. Molec. Biol. 46:51-96.

Cox, D.R. (1962). *Renewal Theory*. Chapman and Hall.

Cox, D.R. and Miller, H.D. (1965). *The Theory of Stochastic Processes*. Chapman and Hall. New York.

Crank, J.C. (1975). *The Mathematics of Diffusion*. Clarendon Press, New York.

Creighton, T.E. (1983). *Proteins: Structures and Molecular Properties*. W.H. Freeman & Co., New York.

Czerminski, R. and R. Elber. (1990). Reaction path study of conformational transitions in flexible systems: applications to peptides. J. Chem. Phys. 92:5580-5601.

Daniel, V.V. (1967). *Dielectric Relaxation*. Academic Press, New York.

Davidson, E.R. (1993). Molecular mechanics and modeling: Overview. Chem. Rev. 93:2337.

Deb, B.M. (1981). *The Force Concept in Chemistry.* Van Nostrand Reinhold, New York.

Doob, J.L. (1953). *Stochastic Processes*. John Wiley & Sons, New York.

Dykstra, C.E. (1993). Electrostatic interaction potentials in molecular force fields. Chem. Rev. 93: 2339-2353.

Eisenberg, R.S. (1993). From structure to permeation in open ionic channels. Biophys. J. 64:A22.

Eisenberg, R.S. (1990). Channels as enzymes. J. Memb. Biol. 115: 1-12.

Eisenberg, R.S., Klosek, M.M. and Schuss, Z. (1994). Stochastic trajectories of diffusion between two concentrations: A model of an open ionic channel. *Submitted.*

Elber, R., Chen, D., Rojewska, D. and Eisenberg, R.S. (1994). Sodium in gramicidin: An example of a permion. *Submitted*.

Eringen, A.C. and Maugin, G.A. (1990). ***Electrodynamics of Continua I. Foundations and Solid Media.*** Springer Verlag, New York.

Fersht, A. (1985). ***Enzyme Structure and Mechanism***. W.H. Freeman & Co., New York.

Forbush, B. (1987). Rapid release of $^{42}$K or $^{86}$Rb from two distinct transport sites on the Na, K-pump in the presence of $P_i$ or vanadate. J. Biol. Chem. 262: 11116-11127.

Franks, F. (1973). ***Water***. Plenum Press, New York.

Friedman, H.L. (1962). ***Ionic Solution Theory***. Interscience Publishers, New York.

Fröhlich, H. (1958). ***Theory of Dielectrics***., (2nd Edition), Oxford University Press, London.

Gard, T.C. (1988). ***Introduction to Stochastic Differential Equations.*** Marcel Dekker, Inc., New York.

Gardiner, C.W. (1985). *Handbook of Stochastic Methods: For Physics, Chemistry and the Natural Sciences*. Springer-Verlag, New York.

Gardner, D. (1992). A time integral of membrane currents. Physiological Reviews. 72:S1-4.

Gates, P.Y., Cooper, K.E. and Eisenberg, R.S. (1990). Analytic diffusion models for membrane channels. *In*: *Ion Channels*. Volume 2. Edited by T. Narahashi, Plenum Press, p. 233-281.

Ghausi, M.S. and Kelly, J.J. (1968). *Introduction to Distributed-Parameter Networks*. Holt, Reinhart and Winston, New York.

Ghze, R. (1988). *A Primer of Diffusion Problems*. John Wiley & Sons, New York.

Gilson, M.K. and Honig, B. (1988). Calculation of the total electrostatic energy of a macromolecular system. *In*: *Conformations and Forces in Protein Folding*. B.T. Nall and K.A. Dill (eds). American Association for the Advancement of Science, Washington, pp. 1-17.

Grant, E.H., R.J. Sheppard and G.P. South. (1978). *Dielectric Behavior of Biological Molecules in Solution*. Oxford University Press, London.

Grabiner, J. (1981). *The Origins of Cauchy's Rigorous Calculus*. MIT Press, Cambridge MA.

Gray, A. (1908). *Lord Kelvin.* (Reprinted 1973, Chelsea Publishing, N.Y.)

Gray, C.G. and Gubbins, K.E. (1984). *Theory of Molecular Fluids*. Vol. 1: Fundamentals. Clarendon Press, Oxford.

Gut, A. (1988). ***Stopped Random Walks***. Springer Verlag, New York.

Haile, J.M. (1992). ***Molecular Dynamics Simulation***. John Wiley & Sons, New York.

Hainsworth, A.H., Levis, R.A. and Eisenberg, R.S. (1994). Origins of open-channel noise in the large potassium channel of sarcoplasmic reticulum. *Submitted.*

Hänggi, P., Talkner, P. and Borokovec, M. (1990). Reaction-rate theory: fifty years after Kramers. Reviews of Modern Physics 62: 251-341.

Hansen, J.P. and McDonald, I.R. (1986). ***Theory of Simple Liquids***. (2nd Edition), Academic Press, New York.

Harned, H.S. and Owen, B.B. (1958). ***The Physical Chemistry of Electrolytic Solutions*** , (3rd Edition), Reinhold Publishing Corporation, New York.

Hedvig, P. (1977). ***Dielectric Spectroscopy of Polymers***. John Wiley & Sons, New York.

Hess, K. (1988). ***Advanced Theory of Semiconductor Devices***. Prentice Hall, New Jersey.

Hill, T.L. (1956). ***Statistical Mechanics***. Dover Publications, New York.

Hill, T.L. (1977). ***Free Energy Transduction in Biology***. Academic Press, New York.

Hill, T.L. (1985). ***Cooperativity Theory in Biochemistry***. Springer-Verlag, New York.

Hille, B. (1992). *Ionic Channels of Excitable Membranes*. 2nd Edition, Sinauer Associates Inc., MA, pp. 1-607.

Hille, B. (1989). Transport Across Cell Membranes: Carrier Mechanisms, Chapter 2, *In: Textbook of Physiology, Vol. 1,* 21st Edition, Edited by Patton, H.D. Fuchs, A.F., Hille, B., Scher, A.M. and Steiner, R.D. WB Saunders Company, Philadelphia. pp. 24-47.

Hirschfelder, J.O. and Curtis, C.F. (1954). *Molecular Theory of Gases and Liquids.* John Wiley & Sons, New York.

Hobson, E.W. (1965). *The Theory of Spherical and Ellipsoidal Harmonics*. Chelsea Publishing Company, New York.

Hodgkin, A.L. (1951). The ionic basis of electrical activity in nerve and muscle. Biol. Rev. 26: 339-409.

Hodgkin, A.L., and Huxley, A.F. (1952). A quantitative description of membrane current and its application to conduction and excitation in nerve. J. Physiol. (London). 117: 500-544.

Hodgkin, A.L. and Keynes, R.D. (1955*a*). Active transport of cations in giant axons from *Sepia* and *Loligo*. J. Physiol. (London) 128:28-60.

Hodgkin, A.L. and Keynes, R.D. (1955*b*). The potassium permeability of a giant nerve fibre. J. Physiol. (London) 128: 61-88.

Hoffman, J.F. and Forbush, B. (1983). Current topics in membranes and transport. *In*: *Structure, Mechanism, and Function of the Na/K Pump*. Vol. 19. Academic Press, New York.

Hohenberg, P. and Kohn, W. (1964). Inhomogeneous electron gas. Phys. Rev. 136 B864.

Holonyak, N. (1992). John Bardeen and the point-contact transistor. Physics Today, 45:36-43.

Hoover, W.G. (1991). *Computational Statistical Mechanics.* Elsevier, New York.

Horowitz, P. and Hill, W. (1989). *The Art of Electronics*. (2nd Edition), Cambridge University
        Press, Cambridge.

Hynes, J.T. (1986). The theory of reactions in solution. *In*: *Theory of Chemical Reactions*.
        M. Baer, (ed), CRC Press, Boca Raton, Fl.

Israelachvili, J.N. (1985). *Intermolecular and Surface Forces*. Academic Press, New York.

Jack, J.J.B., Noble, D. and Tsien, R.W. (1975). *Electric Current Flow in Excitable Cells*.
        Oxford, Clarendon Press, New York.

Jackson, J.D. (1975). *Classical Electrodynamics*. 2nd Edition, Wiley, New York.

Jacobs, M.H. (1935). *Diffusion Processes*. Springer-Verlag, New York.

Jacquez, J.A. (1985). *Compartmental Analysis in Biology and Medicine*. University of Michigan
        Press, Ann Arbor, MI.

Jakobsson, E. (1993). Hierarchies of simulation methods in understanding biomolecular function.
        Int. J. Quantum Chemistry: Quantum Biology Symposium 20:25-36.

Jakobsson, E. and Chiu, S.W. (1987). Stochastic theory of ion movement in channels with single-ion occupancy. Biophys. J. 52: 33-45.

Jakobsson, E. and Chiu, S.W. (1988). Application of Brownian motion theory to the analysis of membrane channel ionic trajectories calculated by molecular dynamics. Biophys. J. 54: 751-756.

Jerome, J.W. (1994). *Mathematical Theory and Approximation of Semiconductor Models*. Springer-Verlag, New York.

Karlin, S. and Taylor, H.M. (1975). *A First Course in Stochastic Processes*. Academic Press, New York.

Katz, B. (1969). *The Release of Neural Transmitter Substances*. C. Thomas, Springfield, Illinois.

Kelvin, L. (1856). On the theory of the electric telegraph. Phil. Mag. 11: 146-160.

Kelvin, L. (1855). On the theory of the electric telegraph. Proc. R. Soc. 7: 382-399.

Kevorkian, J. and Cole, J.D. (1981). *Perturbation Methods in Applied Mathematics.* Springer-Verlag, New York.

Kevorkian, J. (1990). *Partial Differential Equations*. Wadsworth & Brooks/Cole Advanced Books, California.

Keynes, R.D. and Rojas, E. (1974). Kinetics and steady-state properties of the charged system controlling sodium conductance in the squid giant axon. J.Physiol. (London) 239: 393-434.

King, R.W.P. (1965). *Transmission -Line Theory*. Dover Publications, New York.

Kittel, C. (1976). *Introduction to Solid State Physics*, (5th Edition), John Wiley & Sons, New York.

Kortüm, G. (1965). *Treatise on Electrochemistry*, (2nd Edition), Elsevier, New York.

Kuhn, T.S. (1970). *The Structure of Scientific Revolutions*. (2nd Edition), The University of Chicago Press, Chicago.

Kumpf, R.A., and D.A. Dougherty. (1993). A mechanism for ion selectivity in potassium channels: computational studies of cation-$\pi$ interactions. Science 261:1708-1710.

Läuger, P. (1991). *Electrogenic Ion Pumps*. Sinauer Associates, Sunderland, MA.

Lerche, H.R. (1986). *Boundary crossing of Brownian motion*. Springer Verlag, New York.

Macdonald, J.R. (1987). *Impedance Spectroscopy*. John Wiley & Sons, New York.

McCammon, J.A. and Harvey, S.C. (1987). *Dynamics of Proteins and Nucleic Acids.* Cambridge, New York.

McCourt, F.R.W., Beenakker, J.J.M., Kohler, W.E. and Kuscer, I. (1990). *Nonequilibrium Phenomena in Polyatomic Gases. Dilute Gases*. . Vol. 1 Oxford Science Publications.

Mahan, G.D. and Subbaswamy, K.R. (1990). ***Local Density Theory of Polarizability***. Plenum Press, New York.

Maitland, G.C., Rigby, M., Smith, E.B. and Wakeham, W.A. (1987). ***Intermolecular Forces***. Oxford Science Publications, Oxford.

Markowich, P.A. (1986). ***The Stationary Semiconductor Device Equations***. Springer Verlag, New York.

Martynov, G.A. (1992). ***Fundamental Theory of Liquids***. Adam Hilger, New York.

Matsuda, H. (1988). Open-state substructure of inwardly rectifying potassium channels revealed by magnesium block in guinea-pig heart cells. J. Physiol. (London) 397: 237-258.

Mayer, J.E. and Mayer, M.G. (1940). ***Statistical Mechanics***. John Wiley & Sons, New York.

Miller, C. (1982). *bis*-quaternary ammonium blockers as structural probes of the sarcoplasmic reticulum K+ channel. J. Gen. Physiol. 79: 869-891.

Mock, M.S. (1983). ***Analysis of Mathematical Models of Semiconductor Devices.*** Boole Press, Dublin.

Moon, F.C. (1987). ***Chaotic Vibrations***. John Wiley & Sons, New York.

Naeh, T., Klosek, M.M., Matkowsky, B.J. and Schuss, Z. (1990). A direct approach to the exit problem. Siam J. Appl. Math 50: 595-627.

Nauman, E.B. and Buffman, B.A. (1983). ***Mixing in Continuous Flow Systems***. John Wiley & Sons, New York.

Nitzan, A. (1988). Activated rate processes in condensed phases: The Kramers theory revisited. *In*: ***Evolution of Size Effects in Chemical Dynamics. Part 2.*** I. Prigogine, S.A. Rice (eds). John Wiley & Sons, New York, pp. 489-555.

Nitzan, A. and Ratner, M.A. (1994). Conduction in polymers: dynamic disorder transport. J. Phys. Chem. 98: 1765-1775.

Noda, M., Shimizu, S., Tanabe, T., Takai, T., Kayano, T., Ikeda, T., Takahashi, H., Nakayama, H., Kanaoka, Y., Minamino, N., Kangawa, K., Matsuo, H., Raftery, M.A., Hirose, T., Inayama, S., Hayashida, H., Miyata, T. and Numa, S. (1984). Primary structure of *Electrophorus electricus* sodium channel deduced from cDNA sequence. Nature (London) 312: 121-127.

Nowak, W, Czerminski, R. and Elber, R. (1991). Reaction path study of ligand diffusion in proteins: application of the self-avoiding walk (SPW) method to calculate reaction coordinates for the motion of CO through hemoglobin. J.Amer.Chem. Soc. 113:5627-5637.

Onsager, L. and Fuoss, R.M. (1932). Irreversible processes in electrolytes. J. Phys. Chem. 36: 2689-2778.

Panofsky, W. and Phillips, M. (1962). ***Classical Electricity and Magnetism***. Addison Wesley, 2nd edition, pp. 1-494.

Parr, R.G. and Yang, W. (1989). ***Density-Functional Theory of Atoms and Molecules.*** Oxford.

Pesce, A. (1994). Razor blade of life. Nature 368:93.

Pine, J. (1992) An extraordinary man: reflections on John Bardeen. Physics Today 45: 64-70.

Pohl, H.A. (1978). *Dielectrophoresis*. Cambridge University Press, New York.

Purcell, E.M. (1977). Life at low Reynolds number. Amer. J. Phys. 45: 3-11.

Richard, E.A. and Miller, C. (1990). Steady-state coupling of ion channel conformations to a transmembrane ion gradient. Science 247: 1208-1210.

Rigby, M., Smith, E.B., Wakeham, W.A, and Maitland, G.C. (1986). *The Forces Between Molecules*. Oxford Science Publications, London.

Risken, H. (1984). *The Fokker-Planck Equation: Methods of Solution and Applications.* Springer-Verlag, New York.

Robinson, R.A. and Stokes, R.H. (1959). *Electrolyte Solutions,* (2nd Edition), Butterworths Scientific Publications, London.

Rogers, N.K. (1990). The role of electrostatic interactions in the structure of globular proteins. *In*: *Prediction of Protein Structure and The Principles of Protein Conformation*. G.D. Fasman (ed), Plenum Press, New York, pp. 359-389.

Romano, J.P. and Siegel, A.F. (1986). *Counterexamples in Probability and Statistics*. Wadsworth & Brooks/Cole Advanced Books & Software, Monterey, California.

Roulston, D.J. (1990). *Bipolar Semiconductor Devices*. McGraw-Hill Publishing Company, New York.

Rowlinson, J.S. and Swinton, F.L. (1982). *Liquids and Liquid Mixtures.* Butterworth Scientific. New York.

Sakmann, B. and Neher, E. (1983). *Single Channel Recording*. Plenum Press, New York.

Scheraga, H. A. (1968). Calculations of conformations of polypeptides. Advances in Physical Organic Chemistry. 6: 103.

Schneider, M.F., and Chandler, W.K. (1973). Voltage dependent charge movement in skeletal muscle: a possible step in excitation-contraction coupling. Nature 242: 244-246.

Schultz, S.G. (1980). *Basic Principles of Membrane Transport*. Cambridge University Press, New York.

Schultz, G.E. and Schirmer, R.H. (1979). *Principles of Protein Structure*. Springer-Verlag, New York.

Schuss, Z. (1980). *Theory and Applications of Stochastic Differential Equations*. John Wiley & Sons, New York.

Scott, S.K. (1991). *Chemical Chaos*. Clarendon Press, Oxford.

Seeger, K. (1991). *Semiconductor Physics,* (5th Edition), Springer Verlag, New York.

Shockley, W. (1950). ***Electrons and Holes in Semiconductors with Applications to Transistors Electronics.*** D. Van Nostrand Co., Inc., New York.

Shur, M. (1990). ***Physics of Semiconductor Devices***. Prentice Hall, New Jersey.

Smith, C, and Wise, M.N. (1989). ***Energy and Empire. A biographical sketch of Lord Kelvin.*** Cambridge University Press, N.Y.

Smith, T.G., Lecar, H., Redman, S.J. and Gage, P.W. (1985). ***Voltage and Patch Clamping With Microelectrodes***. American Physiological Society, Bethesda, Maryland.

Soukhanov, A.H. (1992). ***The American Heritage Dictionary of the English Language***, (3rd Edition), Houghton Mifflin Company, New York.

Steinfeld, J.I., Francisco, J.S. and Hase, W.L. (1989). ***Chemical Kinetics and Dynamics***. Prentice Hall, New Jersey.

Sten-Knudsen, O. (1978). Passive transport processes. *In*: ***Membrane Transport in Biology: Concepts and Models, Vol. 1***. Ed. D.C. Tosteson. Springer-Verlag, New York, pp. 5-113.

Stigler, S.M. (1986). ***The History of Statistics.*** Harvard University Press, Cambridge MA.

Stratton, J.A. (1941). ***Electromagnetic Theory***. McGraw-Hill Publishing, New York, p. 1-615.

Sze, S.M. (1981). ***Physics of Semiconductor Devices***. John Wiley & Sons, New York.

Szekely, G.J. (1986). ***Paradoxes in Probability Theory and Mathematical Statistics***. D. Reidel Publishing Co.

Taylor, P.D. (1987). ***Thyristor Design and Realization***. John Wiley & Sons, New York.

Tyrrell, H.J.V. and Harris, K.R. 1984. ***Diffusion in Liquids***. Butterworths, Boston.

Ussing, H.H. (1949). The distinction by means of tracers between active transport and diffusion. Acta Physiol. Scand. 19: 43-56.

Walsh, C. (1979). ***Enzymatic Reaction Mechanisms***. W.H. Freeman Company, San Francisco.

Ward, C.A (1977). The rate of gas absorption at a liquid interface. J. Chem. Phys. 67: 229-239.

Ward, C.A. (1983). Effect of concentration on the rate of chemical reactions. J. Chem Phys. 79: 5605-5615.

Ward, C.A. and Elmoselhi, M. (1986). Molecular adsorption at a well defined gas-solid interphase: Statistical rate theory approach. Surface Science 176: 457-475.

Weiss, M.S., Abele, U., Weckesser, J., Welte, W., Schiltz, E. and Schulz, G.E. (1991). Molecular architecture and electrostatic properties of a bacterial porin. Science 254: 1627-1630.

Weiss, M.S., Wacker, T., Weckesser, J., Welte, W. and Schulz, G.E. (1990). The three dimensional structure of porin from *Rhodobacter capsulatus* at 3 Å resolution. FEBS Letters 267: 268-272.

Zauderer, E. (1983). ***Partial Differential Equations of Applied Mathematics***. John Wiley & Sons.